\documentclass[journal]{IEEEtran}
\usepackage{cite}

\ifCLASSINFOpdf
\else
\fi
\usepackage{amsmath}
\ifCLASSOPTIONcompsoc
  \usepackage[caption=false,font=normalsize,labelfont=sf,textfont=sf]{subfig}
\else
  \usepackage[caption=false,font=footnotesize]{subfig}
\fi

\usepackage{amssymb}
\usepackage{booktabs}
\usepackage{multirow}
\usepackage[hidelinks]{hyperref}
\usepackage{siunitx}
\usepackage{mathtools,mathrsfs}
\usepackage{comment}
\usepackage{color}
\usepackage{threeparttable}
\usepackage{bbm}

\def\equationautorefname~#1\null{(#1\null)}

\usepackage{xspace}
\makeatletter
\DeclareRobustCommand\onedot{\futurelet\@let@token\@onedot}
\def\@onedot{\ifx\@let@token.\else.\null\fi\xspace}
\def\eg{e.g\onedot} 
\def\ie{i.e\onedot} 
\def\cf{cf\onedot} 
\def\etc{etc\onedot} 
 
\def\etal{et al\onedot}
\makeatother

\newcommand{\vect}[1]{\mbox{\boldmath $#1$}}

\newcommand{\norm}[1]{\left\lVert#1\right\rVert}

\newcommand{\trans}[1]{#1^\mathsf{T}}

\DeclareMathOperator*{\argmax}{arg\,max}
\DeclareMathOperator*{\argmin}{arg\,min}\def\appendixautorefname~#1\null{~#1 \null}
\def\diag{\mathop{\rm diag}\nolimits}

\makeatletter
\newcommand{\figcaption}[1]{\def\@captype{figure}\caption{#1}}
\newcommand{\tblcaption}[1]{\def\@captype{table}\caption{#1}}
\makeatother

\begin{document}
%
\title{Online Neural Diarization of Unlimited Numbers of Speakers Using Global and Local Attractors}
%
%
%

\author{Shota Horiguchi,~\IEEEmembership{Member,~IEEE,}
        Shinji Watanabe,~\IEEEmembership{Fellow,~IEEE,}
        Paola Garc\'{i}a~\IEEEmembership{Member,~IEEE,}\\
        Yuki Takashima,~\IEEEmembership{}
        Yohei Kawaguchi,~\IEEEmembership{Senior Member,~IEEE,}
\thanks{S. Horiguchi, Y. Takashima, and Y. Kawaguchi are with Hitachi, Ltd.}
\thanks{S. Watanabe is with Carnegie Mellon University.}%
\thanks{P. Garc\'{i}a is with Johns Hopkins University.}%
}

%
%

\markboth{Journal of \LaTeX\ Class Files,~Vol.~14, No.~8, August~2015}%
{Shell \MakeLowercase{\textit{et al.}}: Bare Demo of IEEEtran.cls for IEEE Journals}
%



\maketitle

\begin{abstract}
    A method to perform offline and online speaker diarization for an unlimited number of speakers is described in this paper.
    End-to-end neural diarization (EEND) has achieved overlap-aware speaker diarization by formulating it as a multi-label classification problem. It has also been extended for a flexible number of speakers by introducing speaker-wise attractors.
    However, the output number of speakers of attractor-based EEND is empirically capped; it cannot deal with cases where the number of speakers appearing during inference is higher than that during training because its speaker counting is trained in a fully supervised manner.
    Our method, EEND-GLA, solves this problem by introducing unsupervised clustering into attractor-based EEND.
    In the method, the input audio is first divided into short blocks, then attractor-based diarization is performed for each block, and finally, the results of each block are clustered on the basis of the similarity between locally-calculated attractors.
    While the number of output speakers is limited within each block, the total number of speakers estimated for the entire input can be higher than the limitation.
    To use EEND-GLA in an online manner, our method also extends the speaker-tracing buffer, which was originally proposed to enable online inference of conventional EEND.
    We introduce a block-wise buffer update to make the speaker-tracing buffer compatible with EEND-GLA.
    Finally, to improve online diarization, our method improves the buffer update method and revisits the variable chunk-size training of EEND.
    The experimental results demonstrate that EEND-GLA can perform speaker diarization of an unseen number of speakers in both offline and online inferences.
\end{abstract}


%
\IEEEpeerreviewmaketitle

\section{Introduction}
\IEEEPARstart{I}{dentifying} who spoke when from the input audio is referred to as speaker diarization \cite{anguera2012speaker,park2022review}.
It is a core technology of spoken language understanding of multi-talker conversations in various scenarios such as everyday conversations \cite{kenny2010diarization,barker2018fifth,watanabe2020chime}, doctor-patient conversations \cite{finley2018automated}, meetings \cite{araki2008doa}, lectures \cite{zhu2007multi}, and video contents \cite{bost2015audiovisual}.

While cascaded methods for diarization have been widely investigated in the literature \cite{wang2018speaker,park2020auto,bredin2021endtoend,landini2022bayesian}, progress on end-to-end methods has enabled highly accurate speaker diarization \cite{fujita2019end2,horiguchi2020endtoend,horiguchi2022encoderdecoder,medennikov2020targetspeaker,maiti2021endtoend,liu2021endtoend,zeghidoure2021dive}.
One reason for this advance is the easy handling of overlapping speech.
In cascaded methods, the speaker embeddings extracted for each short segment are clustered to perform diarization.
Therefore, overlap-aware speaker diarization cannot be done unless overlap detection and speaker assignment are performed as post-processing \cite{boakye2008overlapped,landini2020but,horiguchi2021endtoend}.
However, most end-to-end models can naturally handle overlapping speech because they estimate speech segments of multiple speakers simultaneously like multi-label classification.
Another reason is the ease of optimization as an entire diarization system.
In cascaded methods, each module (speech activity detector, speaker embedding extractor, clustering model, \etc) is trained independently, which makes it difficult to optimize the overall diarization system.
In contrast, end-to-end methods use a single neural network to obtain diarization results directly from the input audio, making optimization easier than cascaded methods.
This is also the same for online diarization.
Online diarization with cascaded methods requires all the modules above to enable online use.
On the other hand, an end-to-end method still requires a single model by simply replacing the network architecture with the one that enables online inference \cite{han2021bwedaeend}.
In other methods, an end-to-end model trained for offline use can be used for online purposes by using a buffer to store the previous input-result pairs \cite{xue2021online,xue2021online2}.

Although end-to-end methods have several advantages over cascaded methods, they still have room for improvement.
The biggest challenge is in the estimation of the number of speakers.
In cascaded methods, the number of speakers is estimated as the result of clustering; thus, the number of speakers can be flexibly determined and unlimited.
In contrast, most end-to-end methods fix the number of output speakers due to their network architecture \cite{fujita2019end1,fujita2019end2}.
Most methods that enable the inference of a flexible number of speakers conduct it by outputting null speech activities for absent speakers, so the maximum number of speakers is limited \cite{maiti2021endtoend,he2021targetspeaker}.
Some methods use speaker-wise auto-regressive inference to avoid setting the maximum number of speakers by the network architecture; but in practice, the number of output speakers is still capped by the training dataset \cite{horiguchi2020endtoend,fujita2020neural,takashima2021endtoend,horiguchi2022encoderdecoder}.
A few studies, one of which is the basis for this paper, integrated the end-to-end approach with unsupervised clustering to solve this problem \cite{kinoshita2021integrating,kinoshita2021advances,horiguchi2021towards,kinoshita2022tight}.
The methods showed promising results on various benchmark datasets, but their online inferences have not been investigated in the literature.

In this article, we propose end-to-end neural diarization with global and local attractors (EEND-GLA), which integrates attractor-based EEND (EEND-EDA) \cite{horiguchi2020endtoend,horiguchi2022encoderdecoder} with clustering to conduct speaker diarization without limiting the number of speakers.
In addition to the attractors calculated from the entire recording (\ie, \textit{global attractors}) in the same manner as in EEND-EDA, we also utilize attractors calculated from each short block (\ie, \textit{local attractors}) to obtain block-wise diarization results.
Because the set of speakers and their output order may be different among the blocks, we use clustering to find the appropriate speaker correspondence between the blocks on the basis of the similarities between the local attractors.
Here, we assume that the number of speakers appearing in a short period is low, and so the number of speakers within each block can be limited and fixed with a maximum number. However, the total number of speakers is estimated as the result of clustering; it is no longer limited by the network architecture or training datasets.
To enable online inference of EEND-GLA, we also propose a block-wise speaker-tracing buffer, which extends the original speaker-tracing buffer \cite{xue2021online,xue2021online2} to update the buffer elements in a block-wise manner.
With this modification, we can assume that the number of speakers within each block is limited in the buffer as well because each block stores time-consecutive elements.

This paper is organized on the basis of our previous paper \cite{horiguchi2021towards}, in which the fundamental algorithm of EEND-GLA was presented.
Our contributions that differ from those of the previous paper are summarized as follows.
\begin{itemize}
    \item We propose a block-wise speaker-tracing buffer, which enables the online inference of EEND-GLA.
    \item We improve the speaker-tracing buffer by introducing speaker-balanced sampling probabilities.
    \item We revisit variable chunk-size training to improve online diarization, especially at the very beginning of inference.
    \item We evaluate our method on offline and online diarization settings consistently over various prior studies.
\end{itemize}

The organization of this paper is as follows. \autoref{sec:related_work} reviews offline and online diarization methods in the literature. \autoref{sec:conventional_method} details conventional attractor-based EEND (\autoref{sec:eend_eda}) and speaker-tracing buffer that enables its online inference (\autoref{sec:speaker_tracing_buffer}). \autoref{sec:proposed_method} presents proposed EEND-GLA (\autoref{sec:eend_gla}) and some modifications to the speaker-tracing buffer to make it compatible with EEND-GLA and improve its performance (Sections \ref{sec:bw_stb} to \ref{sec:vct}). Sections \ref{sec:experimental_settings} and \ref{sec:results} describe the experimental settings and results, respectively. \autoref{sec:conclusion} concludes the paper.

\section{Related work}\label{sec:related_work}
\subsection{Offline diarization}\label{related_work_offline_diarization}
The conventional cascaded approach for speaker diarization consists of the following operations: 1) speech activity detection (SAD), 2) speaker embedding extraction from each detected speech segment, 3) clustering of the embeddings, and 4) optional overlap handling.
The oracle SAD is sometimes used in the experiments, but the remaining parts are actively being studied in the literature: better speaker embedding extraction methods \cite{snyder2018xvectors,diez2019bayesian,xiao2021microsoft,dawalatabad2021ecapatdnn}, clustering methods \cite{park2020auto,singh2021selfsupervised,landini2022bayesian}, and overlap assignment methods \cite{landini2020but,bullock2020overlap,jung2021threeclass}.
The cascaded approach is based on unsupervised clustering; thus, the number of output speakers can take an arbitrary value and can be set flexibly during inference.

Different end-to-end approaches for speaker diarization have been studied, but they have drawbacks when performing speaker diarization without any restrictions.
Some methods such as personal VAD \cite{ding2020personal} and VoiceFilter-Lite \cite{wang2020voicefilterlite} are not suited for speaker-independent diarization because they require a target speaker's embedding vector (\eg, d-vector) for inference.
Target-speaker voice activity detection (TS-VAD) accepts multiple speakers' embeddings, but they have to be obtained in advance from another diarization method such as a cascaded-based approach \cite{medennikov2020targetspeaker} or end-to-end neural diarization (EEND) \cite{wang2022incorporating}.
The initial models of TS-VAD and EEND \cite{fujita2019end1,fujita2019end2} fix the output number of speakers with their network architectures, so they are not suited for diarization of an unknown number of speakers.
The recurrent selective attention network (RSAN) \cite{kinoshita2018listening}, or some extensions of TS-VAD \cite{he2021targetspeaker} and EEND \cite{horiguchi2020endtoend,fujita2020neural,maiti2021endtoend} can deal with a flexible number of speakers.
However, the TS-VAD-based methods explicitly determine the maximum number of outputs with their network architecture, except for a few recent attempts such as multi-target filter and detector (MTFAD) \cite{cheng2022multi} and Transformer-based TS-VAD \cite{wang2022target}.
The EEND-based methods do not have such explicit limitations, but the maximum number of speakers is empirically known to be bounded by their training datasets.
Whether or not RSAN can deal with an unlimited number of speakers is unclear because only the speaker counting accuracies on zero, one, and two-speaker conditions were reported in the paper and each was observed during training.
Making the number of output speakers not only flexible but also unlimited is an important challenge for end-to-end diarization.

The combination of an end-to-end approach and clustering is a promising direction to solve the problem of the limitation of the number of speakers.
For example, EEND as post-processing \cite{horiguchi2021endtoend} and overlap-aware resegmentation \cite{bredin2021endtoend} use EEND to refine the results obtained with cascaded diarization systems.
Multi-scale diarization decoder \cite{park2022multiscale} also employs a similar post-processing approach.
In these methods, the initial results are based on the clustering of speaker embeddings; hence, the number of output speakers can be arbitrary.
However, this nullifies the main advantage of the end-to-end approach, that is, simplicity.
The other approach is EEND-vector clustering \cite{kinoshita2021integrating,kinoshita2021advances,kinoshita2022tight}.
It uses EEND for shortly divided blocks and then finds the speaker corresponding between them using speaker embeddings.
It is relevant to our method in this paper, but some differences exist between them.
One is that EEND-vector clustering requires unique speaker identity labels \textit{over} the recordings in the training set.
This means that we must know whether or not a pair of speakers that appeared in different recordings has the same identity.
Such information can be easily obtained from simulation data but is not always suitable for real recordings.
EEND-GLA only requires the speaker labels \textit{within} each recording; thus, we can use such real recordings for training.
This property is also powerful when conducting, for example, unsupervised or semi-supervised domain adaptation \cite{takashima2021semisupervised}.
Another difference is that EEND-vector clustering requires a somewhat long length of blocks (\eg, \SI{30}{\second}) to obtain reliable speaker embeddings to achieve the best performance.
However, because the number of output speakers within a block is limited by the network architecture, the length would result in a limited output number of speakers in the final results.
Another problem is that the length causes a severe latency if we want to use it for online inference.
However, EEND-GLA splits a sequence into short blocks after generating frame-wise embeddings from acoustic features using stacked Transformer encoders.
As a result, the frame-wise embeddings can capture the global context, so we can use a lower block length (\SI{5}{\second} in this paper) than EEND-vector clustering.

\subsection{Online diarization}\label{sec:related_work_online_diarization}
There are also cascaded and end-to-end approaches to online diarization.
In cascaded approaches, of course, all modules have to work in an online manner.
The most crucial part is a clustering of speaker embeddings, and many methods have been proposed for that in the literature, \eg, UIS-RNN \cite{zhang2019fully}, UIS-RNN-SML \cite{fini2020supervised}, constraint incremental clustering in overlap-aware online speaker diarization \cite{coria2021overlapaware}, and turn-to-diarize \cite{xia2021turntodiarize}.
Basically, online clustering is not as good as offline clustering.
In particular, VBx \cite{landini2022bayesian}, the current state-of-the-art offline clustering method for diarization, relies on two-stage clustering to refine the results and thus is difficult to be used for online inference.
In fact, even if the rest of the modules are similar between offline and online methods, replacing VBx with online clustering reportedly causes a significant drop in performance \cite{coria2021overlapaware}.

On the other hand, end-to-end approaches have also been explored in online diarization.
Online diarization with end-to-end models has two directions.
One is to train a model with frame-wise or block-wise inputs separately from the offline model.
For example, Online RSAN \cite{von2019all,kinoshita2020tackling} is trained with block-wise inputs to extend the original RSAN \cite{kinoshita2018listening} for an online manner. This method uses speaker embeddings to convey information between blocks to make the order of output speakers consistent.
BW-EDA-EEND \cite{han2021bwedaeend} replaced the Transformer encoders in EEND with Transformer-XL \cite{dai2019transformerxl} to extend EEND-EDA \cite{horiguchi2020endtoend,horiguchi2022encoderdecoder} to deal with block-wise inputs.
In this method, the hidden state embeddings obtained during the processing of previous blocks are used to process the current block, thereby solving the speaker permutation ambiguity between blocks.
This direction is beneficial to optimize online diarization itself, but the training cost is doubled if we need to prepare diarization systems for both offline and online inference independently.
The other possibility is to divert an offline diarization model for online inference.
For this purpose, speaker-tracing buffer \cite{xue2021online,xue2021online2} has been proposed to implement online inference of EEND with no modification of the network architecture.
It stores acoustic features and their corresponding diarization results of the selected past frames to solve the speaker permutation ambiguity (see \autoref{sec:speaker_tracing_buffer} for a detailed explanation).
Because it was reported that EEND-EDA with speaker-tracing buffer outperformed BW-EDA-EEND \cite{xue2021online2}, we focused on this direction in this study.

\section{Conventional method}\label{sec:conventional_method}
\subsection{Attractor-based end-to-end neural diarization}\label{sec:eend_eda}
End-to-end neural diarization (EEND) is a framework to estimate multiple speakers' speech activities from the input audio.
In particular, attractor-based EEND (EEND-EDA) \cite{horiguchi2020endtoend,horiguchi2022encoderdecoder} also estimates the number of speakers simultaneously.
Given $T$-length $F$-dimensional acoustic features $X\in\mathbb{R}^{F\times T}$,
they are first converted to the same length of $D$-dimensional frame-wise embeddings $E\in\mathbb{R}^{D\times T}$ using stacked Transformer encoders:
\begin{equation}
    E=\mathsf{TransformerEncoder}\left(X\right)\in\mathbb{R}^{D\times T}.\label{eq:encoder}
\end{equation}
Then, the encoder-decoder-based attractor calculation module (EDA) calculates attractors $\vect{a}_s\in\left(0,1\right)^{D}$ for each speaker $s\in\mathbb{N}$ from $E$ in \autoref{eq:encoder} in a sequence-to-sequence manner as
\begin{equation}
    \vect{a}_1,\vect{a}_2,\dots=\mathsf{EDA}\left(E\right).\label{eq:eda}
\end{equation}
The decoder calculation in \autoref{eq:eda} continues as long as the attractor existence probability $\hat{z}_s$ calculated from $\vect{a}_s$ is not less than 0.5, and the largest $s$ that fulfills $\hat{z}_s\geq0.5$ is the estimated number of speakers $\hat{S}$, as follows:
\begin{align}
    \hat{z}_s&=\sigma\left(\mathsf{Linear}\left(\vect{a}_s\right)\right)\in\left(0,1\right),\label{eq:attractor_existence_probability}\\
    \hat{S}&=\min\left\{s\mid s\in\mathbb{Z}_{\geq 0}\wedge \hat{z}_{s+1}<0.5\right\},\label{eq:estimate_speaker}
\end{align}
where $\sigma\left(\cdot\right)$ is the element-wise sigmoid function.
Finally, the estimations of speech activities $\hat{Y}$ are calculated as dot products between the frame-wise embeddings and attractors with their existence probabilities greater than or equal to 0.5:
\begin{equation}
    \hat{Y}=\sigma\left(\trans{\begin{bmatrix}
    \vect{a}_1&\cdots&\vect{a}_{\hat{S}}
    \end{bmatrix}}E\right)\in\left(0,1\right)^{\hat{S}\times T}.\label{eq:posterior}
\end{equation}

During training, the following loss is used for network optimization:
\begin{equation}
    \mathcal{L}_\text{global}=\mathcal{L}_{\text{diar}}+\alpha\mathcal{L}_{\text{exist}},
    \label{eq:loss_global}
\end{equation}
where $\alpha$ is the weighting parameter, which was set to $1$ in this study.
The first term $\mathcal{L}_{\text{diar}}$ is the permutation-free diarization loss, which optimizes the output speech activities, defined as
\begin{equation}
    \mathcal{L}_\text{diar}=\frac{1}{TS}\argmin_{\phi\in\Phi\left(S\right)}H\left(Y,P_\phi \hat{Y}\right),\label{eq:loss_diar}
\end{equation}
where $\Phi\left(S\right)$ is a set of all the possible permutations of $\left(1,\dots,S\right)$, $P_\phi\in\left\{0,1\right\}^{S\times S}$ is the permutation matrix that corresponds to the permutation $\phi$, $H\left(\cdot,\cdot\right)$ is the sum of element-wise binary cross entropy, and $S$ is the ground-truth number of speakers.
Note that the estimation of speech activities $\hat{Y}$ is calculated using the ground-truth number of speakers during training, \ie, $\hat{Y}\in\left(0,1\right)^{S\times T}$. 
The second term $\mathcal{L}_{\text{exist}}$ is the attractor existence loss, which optimizes the number of output attractors, defined as
\begin{equation}
    \mathcal{L}_\text{exist}=\frac{1}{S+1}\sum_{s=1}^{S+1}H\left(z_{s},\hat{z}_{s}\right),\quad
    z_s=\begin{cases}
        1&(s\in\{1,\dots,S\})\\
        0&(s=S+1)
    \end{cases}.
    \label{eq:loss_att}
\end{equation}
Following the previous study \cite{horiguchi2022encoderdecoder}, the attractor existence loss is used to update only the parameters of the linear layer in \autoref{eq:attractor_existence_probability}.

\subsection{Online diarization with speaker-tracing buffer}\label{sec:speaker_tracing_buffer}
\begin{figure*}[t]
    \centering
    \includegraphics[width=1.0\linewidth]{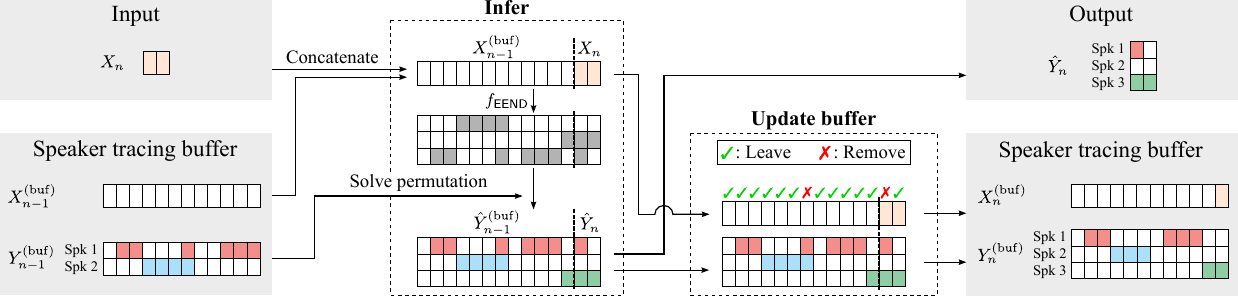}
    \caption{Online diarization using speaker-tracing buffer proposed in \cite{xue2021online,xue2021online2}.}
    \label{fig:framewise_stb}
\end{figure*}
A speaker-tracing buffer has been proposed to enable online inference of EEND without additional training \cite{xue2021online,xue2021online2}.
The speaker-tracing buffer stores the past acoustic features and the corresponding estimation to solve the speaker permutation ambiguity.
The schematic diagram of online diarization using a speaker-tracing buffer is shown in \autoref{fig:framewise_stb}.

In the situation of online diarization, chunked acoustic features sequentially arrive, and the length of each chunk is $\nu$.
Suppose $X_{n-1}^\text{(buf)}\in\mathbb{R}^{F\times T_{n-1}^\text{(buf)}}$ and $Y_{n-1}^\text{(buf)}\in\mathbb{R}^{\hat{S}_{n-1}\times T_{n-1}^\text{(buf)}}$ are features and the corresponding estimations stored in the buffer just before the $n$-th input, respectively, where $T_{n-1}^\text{(buf)}$ is their length and $\hat{S}_{n-1}$ is the previously estimated number of speakers.
Given $n$-th input $X_n\in\mathbb{R}^{F\times\nu}$, it is concatenated with the features in the buffer and processed by EEND $f_\mathsf{EEND}$ as
\begin{equation}
    \begin{bmatrix}\hat{Y}_{n-1}^\text{(buf)}&\hat{Y}_n\end{bmatrix}=f_\textsf{EEND}\left(\begin{bmatrix}X_{n-1}^\text{(buf)}&X_n\end{bmatrix}\right)\in\left(0,1\right)^{\hat{S}'_n\times\left(T_{n-1}^\text{(buf)}+\nu\right)},
\end{equation}
where $\hat{S}'_n$ is the newly estimated number of speakers\footnote{Ideally, $\hat{S}'_n$ is not less than $\hat{S}_{n-1}$.}, and $\hat{Y}_{n-1}^\text{(buf)}$ and $\hat{Y}_n$ are the estimated results that correspond to $X_{n-1}^\text{(buf)}$ and $X_n$, respectively.
Here, the previously estimated number of speakers $\hat{S}_{n-1}$ and the newly estimated one $\hat{S}'_{n}$ may differ, \eg, $\hat{S}_{n-1}=2$ and $\hat{S}'_{n}=3$ in \autoref{fig:framewise_stb}.
To align them to the same number, we first update each of $\hat{Y}_{n-1}^\text{(buf)}\in\left(0,1\right)^{\hat{S}'_n\times T^\text{(buf)}_{n-1}}$ and $\hat{Y}_n\in\left(0,1\right)^{\hat{S}'_n\times \nu}$ to have $\hat{S}_n=\max\left({\hat{S}_{n-1},\hat{S}'_n}\right)$ rows by zero padding.
The order of speakers is then permuted to be aligned to that of $\hat{Y}_{n-1}^\text{(buf)}$ as
\begin{align}
    \begin{bmatrix}\hat{Y}_{n-1}^\text{(buf)}&\hat{Y}_n\end{bmatrix}&\leftarrow P_\psi \begin{bmatrix}\hat{Y}_{n-1}^\text{(buf)}&\hat{Y}_n\end{bmatrix},\label{eq:solve_permutation}\\
    \psi&=\argmax_{\phi\in\Phi\left(\hat{S}_n\right)}\left\langle{Y_{n-1}^\text{(buf)}, P_\phi\hat{Y}_{n-1}^\text{(buf)}}\right\rangle_\mathrm{F},\label{eq:permutation}
\end{align}%
where $\left\langle A,B\right\rangle_\mathrm{F}$ denotes the Frobenius inner product between real-valued two matrices $A=\left[a_{ij}\right]$ and $B=\left[b_{ij}\right]$ defined as\footnote
{In the original STB paper, mean normalization is applied for each of $A$ and $B$ before the calculation of the Frobenius inner product, but it does not affect the result so we omit it here.}
\begin{equation}
    \left\langle A,B\right\rangle_\mathrm{F}\coloneqq\sum_{i,j}a_{ij}b_{ij}.\label{eq:frobenius_ip}
\end{equation}
Note that \autoref{eq:permutation} is executable in polynomial time by using the well-known Hungarian algorithm.
Finally, the permuted $\hat{Y}_n$ is output as the estimated result for $X_n$.

For the next (\ie, $\left(n+1\right)$-th) input, the buffer is updated with the current input features and the corresponding results.
If the buffer length $M$ is large enough to store all the features and results, \ie, $T_{n-1}^\text{(buf)}+\nu\leq M$, we update the buffer using
\begin{align}
    X_n^\text{(buf)}&\leftarrow\begin{bmatrix}X_{n-1}^\text{(buf)}&X_n\end{bmatrix},\\
    Y_n^\text{(buf)}&\leftarrow\begin{bmatrix}\hat{Y}_{n-1}^\text{(buf)}&\hat{Y}_n\end{bmatrix}.
\end{align}
If $T_{n-1}^\text{(buf)}+\nu>M$, only $M$ frames among them are selected to be stored.
The original speaker-tracing buffer mainly utilized the following two update strategies.
\begin{enumerate}
    \item \textbf{First-in-first-out (FIFO)}: acoustic features and results of the latest $M$ frames are always stored in the buffer. Speakers who do not appear in the last $M$ frames are not tracked with this strategy; thus, this strategy alone is not preferable.
    \item \textbf{Sampling}: the features and results of informative $M$ frames to solve speaker permutation ambiguity are selected among $T_{n-1}^\text{(buf)}+\nu$ and stored. In the previous studies \cite{xue2021online,xue2021online2}, sampling probabilities based on Kullback-Leibler (KL) divergence were used. The KL divergence at the $t$-th frame $\omega_t$ is calculated from the speaker-normalized posteriors $\bar{y}_{s,t}$ and the discrete uniform distribution with the posterior probability of $\frac{1}{S_n}$ as
    \begin{align}
        \omega_{t}&=\sum_{s=1}^{S_n}\bar{y}_{s,t}\log \left(\bar{y}_{s,t}S_n\right),\label{eq:conventional_score}\\
        \bar{y}_{s,t}&=\frac{y_{s,t}}{\sum_{s'=1}^{S_n} y_{s',t}}.
    \end{align}
    The sampling probabilities $\tilde{\omega}_t$ are defined as the normalized KL divergence so that the sum is one:
    \begin{equation}
        \tilde{\omega}_t=\frac{\omega_t}{\sum_{t'}\omega_{t'}}.\label{eq:conventional_score_norm}
    \end{equation}
\end{enumerate}

With the aforementioned speaker-tracing buffer, a trained EEND model can be used for online inference as it is.
However, EEND is generally trained with a fixed length of chunks, \eg, 500 frames, so the diarization performance decreases at the very beginning of the inference where the number of frames is low.
To cope with this problem, variable chunk-size training (VCT) was proposed \cite{xue2021online}.
For VCT, the length of each chunk is varied by masking the input minibatch.
It has been evaluated in two-speaker conditions \cite{xue2021online} but has not been evaluated in the flexible-number-of-speaker conditions \cite{xue2021online2}.
Even the two-speaker experiments have a limited analysis of how VCT improved the diarization error rates (DERs).
These aspects will be analyzed in \autoref{sec:eval_stb}.

\section{Proposed method}\label{sec:proposed_method}
\subsection{EEND with global and local attractors}\label{sec:eend_gla}
While EEND-EDA can treat a flexible number of speakers, the maximum number of speakers to be output was empirically revealed to be limited by the dataset used during training.
For example, if EEND-EDA is trained using mixtures, each of which contains at most four speakers, it cannot produce a valid result for the fifth or later speaker even if a mixture contains more than four speakers.

\begin{figure}[t]
    \centering
    \begin{minipage}{\linewidth}
        \centering
        \includegraphics[width=0.32\linewidth]{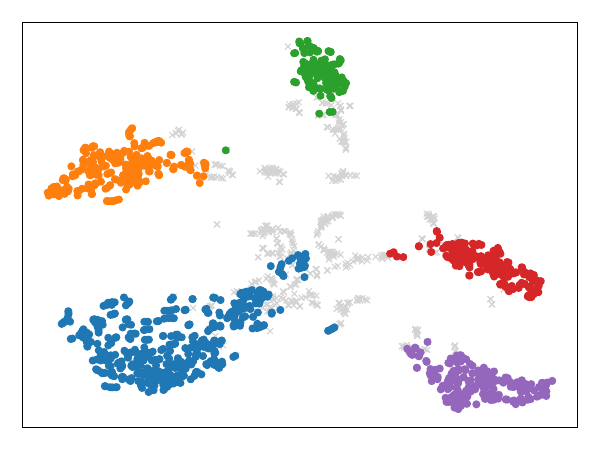}\hfill
        \includegraphics[width=0.32\linewidth]{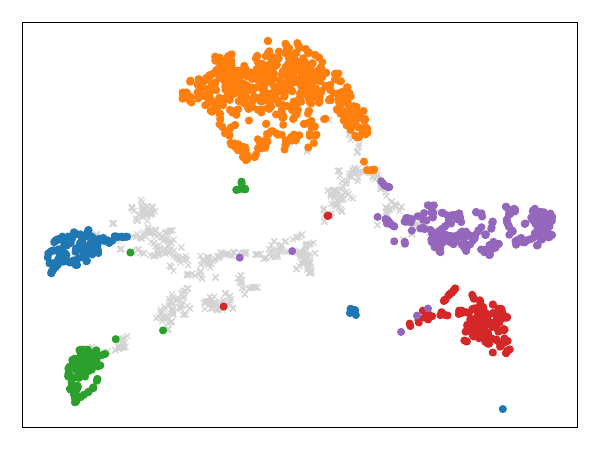}\hfill
        \includegraphics[width=0.32\linewidth]{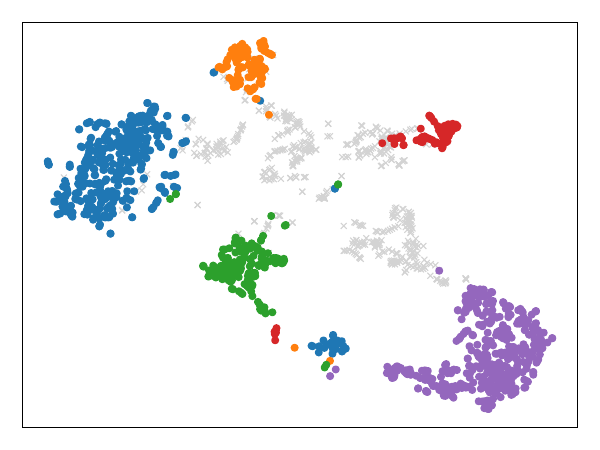}\\
        \includegraphics[width=0.32\linewidth]{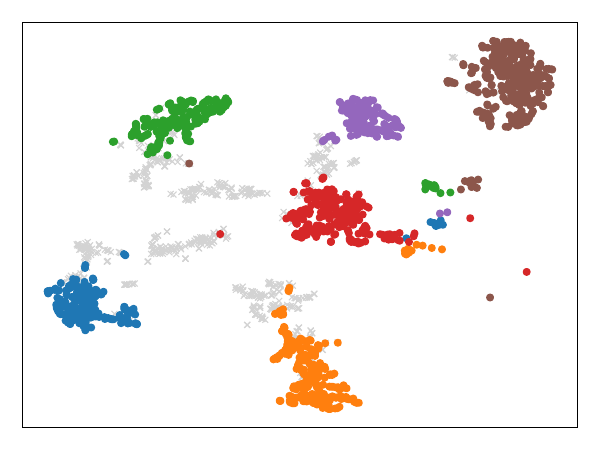}\hfill
        \includegraphics[width=0.32\linewidth]{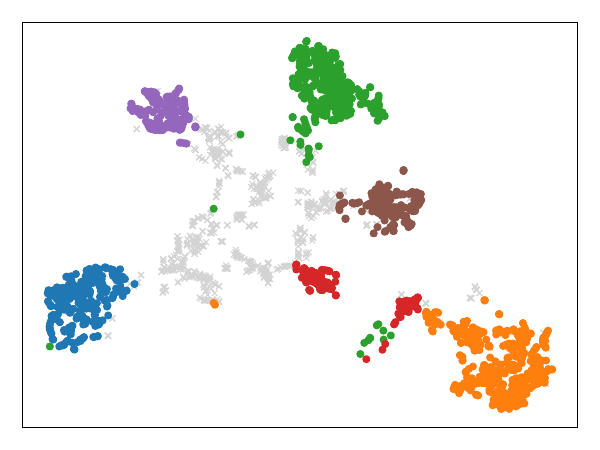}\hfill
        \includegraphics[width=0.32\linewidth]{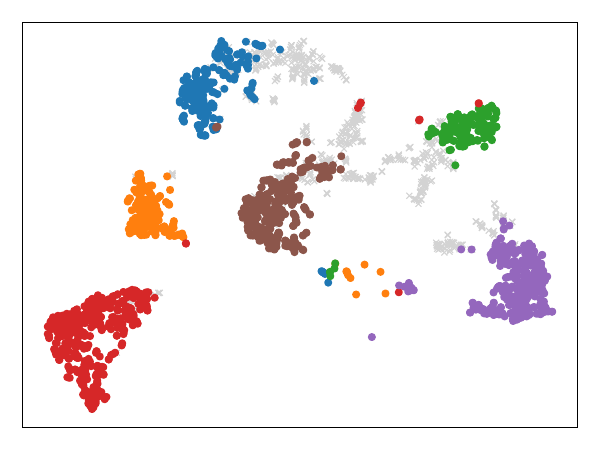}\\
        \caption{t-SNE visualization of frame-wise embeddings extracted from simulated 5-speaker mixtures (top) and 6-speaker mixtures (bottom). The EEND-EDA used for extraction was trained using \{1,2,3,4\}-speaker mixtures. Single-speaker frames are denoted by the dots with colors corresponding to the speaker identities and overlapped frames are denoted by the crosses in light gray. Frames of silence were excluded from the visualization.}
        \label{fig:visualization}
    \end{minipage}
\end{figure}
To reveal which part causes this limitation, we visualized the frame-wise embeddings that were output from the last Transformer encoder using t-SNE \cite{van2008visualizing} in \autoref{fig:visualization}.
Even though EEND-EDA was trained on mixtures, each of which consists of at most four speakers, five or six speakers' speeches were clearly separated in the embedding space.
The visualization revealed that EDA fails to estimate attractors for the unseen-number-of-speaker cases.

The proposed method assumes that the number of speakers that appear in a short period is bounded in practice. We first conduct attractor-based diarization for each short block and then find inter-block speaker correspondence on the basis of the similarity of the attractors.
We call the attractors calculated within each block \textit{local attractors}.
Even if the number of speakers within each block is limited owing to EDA, the total number of speakers within a recording can be higher than the upper bound.
Our method also utilizes global-attractor-based diarization just as EEND-EDA does.

\subsubsection{Training}
\begin{figure*}[t]
    \includegraphics[width=\linewidth]{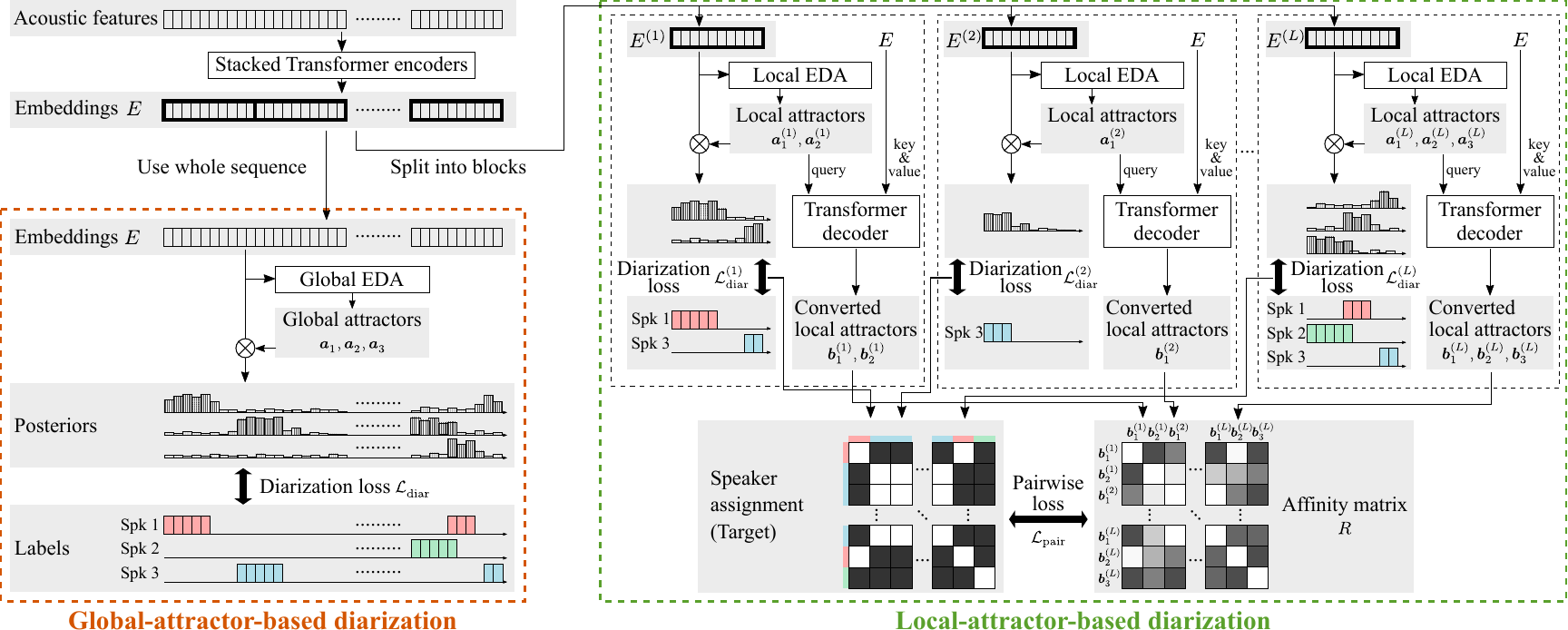}
    \caption{End-to-end neural diarization with global and local attractors (EEND-GLA). The attractor existence losses are omitted from the illustration.}
    \label{fig:eend_eda_gl}
\end{figure*}
\autoref{fig:eend_eda_gl} illustrates the proposed diarization based on global and local attractors, which we call EEND-GLA.
The global-attractor-based diarization is identical to EEND-EDA described in \autoref{sec:eend_eda}; in this section, we introduce local-attractor-based diarization.
Given frame-wise embeddings $E$, we first split them into short blocks, each of which has a length of $\lambda$.
Here, we assume that the sequence of frame-wise embeddings is split into $L$ blocks, \ie, $E\coloneqq\begin{bmatrix}
E^{(1)}&\cdots&E^{(L)}
\end{bmatrix}$, where $E^{(l)}\in\mathbb{R}^{D\times \lambda}$ for $l\in\left\{1,\dots,L\right\}$ and $L\coloneqq\frac{T}{\lambda}$\footnote{
For simplicity, we assume that the length of the sequence $T$ is divisible by $\lambda$, but in practice, the length of the last block can be shorter than $\lambda$.}.
From the $l$-th block, local attractors $\vect{a}_1^{(l)},\dots,\vect{a}_{S_l}^{(l)}\eqqcolon A^{(l)}$ are calculated using \autoref{eq:eda}, and the speech activities for the $l$-th block $\hat{Y}^{(l)}\in\left(0,1\right)^{S_l\times\lambda}$ are calculated using \autoref{eq:posterior}.
Here, $S_l$ is the number of speakers that appeared in the $l$-th block, which satisfies $0\leq S_l\leq S$.
The diarization loss $\mathcal{L}_{\text{diar}}^{(l)}$ and attractor existence loss $\mathcal{L}_{\text{exist}}^{(l)}$ for the $l$-th block are calculated using \autoref{eq:loss_diar} and \autoref{eq:loss_att}, respectively.

The local attractors are clustered to find inter-block speaker correspondence.
Since the local attractors themselves are optimized to minimize the diarization error, non-parametric similarities between them are not fit for speaker clustering, like cascaded methods require a scoring model based on probabilistic linear discriminant analysis. EEND-GLA includes the scoring model equivalent that is jointly optimized with the diarization and attractor existence losses. We first convert them by using the following Transformer decoder:
\begin{equation}
    B^{(l)}=\mathsf{TransformerDecoder}\left(A^{(l)},E,E\right)\in\mathbb{R}^{D\times S_l},
\end{equation}
where the first, second, and third arguments for the Transformer decoder are query, key, and value inputs, respectively.
Here, the converted attractors $B$ are expected to be speaker discriminative within each input audio. Thus, we refer to them as relative speaker embeddings, as contrasted to general speaker embeddings with global discriminability such as x-vectors.
The relative speaker embeddings from all the blocks are gathered $B=\left[\vect{b}_i\right]_i\coloneqq\left[B^{(1)},\dots,B^{(L)}\right]\in\mathbb{R}^{D\times S^*}$ and optimized to minimize the pairwise loss defined as follows:
\begin{align}
    \mathcal{L}_\text{pair}=\sum_{i,j\in\{1,\dots,S^*\}}&\frac{1}{S^2c_ic_j}\left(\vphantom{\left(1-\chi_{ij}\right)\left[\operatorname{sim}\left(\vect{b}_i,\vect{b}_j\right)-\delta\right]_{+}}\chi_{ij}\left(1-\operatorname{sim}\left(\vect{b}_i,\vect{b}_j\right)\right)\right. +\nonumber\\
    &\left.\left(1-\chi_{ij}\right)\left[\operatorname{sim}\left(\vect{b}_i,\vect{b}_j\right)-\delta\right]_{+}\right),\label{eq:pairwise_loss}
\end{align}
\vspace{-10pt}
\begin{equation}
    \chi_{ij}=\begin{cases}
    1&\left(\text{$\vect{b}_i$ and $\vect{b}_j$ correspond to the same speaker}\right)\\
    0&(\text{otherwise})
    \end{cases},
\end{equation}
where $S^*\coloneqq\sum_{l=1}^LS_l$ is the total number of local attractors, $\operatorname{sim}\left(\vect{b}_i,\vect{b}_j\right)\coloneqq\frac{\trans{\vect{b}_i}\vect{b}_j}{\norm{\vect{b}_i}\norm{\vect{b}_j}}$ is the cosine similarity between $\vect{b}_i$ and $\vect{b}_j$, and $\left[\cdot\right]_{+}$ is the hinge function.
$c_i$ ($c_j$) is the number of local attractors that correspond to the $i$-th ($j$-th) attractor's speaker, and this correspondence is obtained by finding the optimal speaker permutation when calculating the diarization loss using \autoref{eq:loss_diar}.
This pairwise loss aims to make the angle between relative speaker embeddings of the same speaker as small as possible and those of different speakers at least $\arccos{\delta}$.
In this paper, we used $\delta=0.5$ during pretraining and $\delta=0$ during adaptation.
Note that this loss definition is based on the contrastive loss used for instance segmentation in computer vision \cite{fathi2017semantic,kong2018recurrent}.
The process of grouping pixel-wise embeddings into instances is very similar to our problem setting of grouping local attractors into speaker identities.
While x-vectors or frame-wise embeddings cannot be hardly assigned to one of the speaker identities because of overlaps, the local attractors can be divided by speaker identities because each of them corresponds to one speaker.

As a result, the loss based on local attractors is defined as
\begin{equation}
    \mathcal{L}_\text{local}=\frac{1}{L}\sum_{l=1}^L\left(\mathcal{L}_\text{diar}^{(l)}+\alpha\mathcal{L}_\text{exist}^{(l)}\right)+\gamma\mathcal{L}_\text{pair},
    \label{eq:local_loss}
\end{equation}
where $\gamma$ is the weighting parameter for which we set $\gamma=1$ in this study.
The total loss of EEND-GLA is defined as a sum of global- and local-attractor-based losses:
\begin{equation}
    \mathcal{L}_\text{both}=\mathcal{L}_\text{local}+\mathcal{L}_\text{global}.
    \label{eq:both_loss}
\end{equation}

\subsubsection{Inference}\label{sec:eend_gla_inference}
During inference, the number of speakers within each block $\hat{S}_l\in\mathbb{Z}_{\geq 0}$ is estimated using \autoref{eq:estimate_speaker}, and speech activities of $\hat{S}_l$ speakers are estimated using \autoref{eq:posterior}.
Speaker correspondence between blocks can be found by clustering the relative speaker embeddings $B$, and the problem here is how to determine the number of clusters.

Some conventional methods based on spectral clustering \cite{wang2018speaker,park2020auto} consist of the following steps: 1) construct an affinity matrix from frame-wise embeddings, 2) calculate its graph Laplacian, 3) use eigenvalue decomposition, and 4) determine the number of speakers as the value that maximizes the eigengap.
Some tricks were used in these studies to reduce the effect of noise in the affinity matrix.
In one study, an affinity matrix calculated from frame-wise d-vectors was smoothed by using Gaussian blur \cite{wang2018speaker}.
Another study utilized $p$ nearest binarization to the affinity matrix to remove unreliable values \cite{park2020auto}.
In our case, local attractors are extracted not only for each block but also for each speaker within a block. Smoothing should be applied along the time axis of each speaker, but in this case, smoothing cannot be performed because the proper inter-block correspondence of the speakers has not been obtained.
In our method, a few local attractors are calculated every five seconds, and hence $p$ nearest neighbor binarization is also not suitable because it generally requires dozens of embeddings per cluster.

Therefore, in EEND-GLA, we use the unprocessed affinity matrix to estimate the number of clusters.
However, if we estimate it based on the eigengaps of graph Laplacian, noises cause a lot of tiny clusters because the size of clusters is not considered in this approach.
Thus, we use the affinity matrix directly instead of its graph Laplacian to penalize small clusters more.
Given the positive-semidefinite affinity matrix $R=\left(r_{ij}\right)\in\left[-1,1\right]^{S^*\times S^*}$, where $r_{ij}=\operatorname{sim}\left(\vect{b}_i,\vect{b}_j\right)$, the number of clusters $\hat{S}$ can be estimated using its eigenratios instead of eigengaps as
\begin{equation}
    \hat{S}=\argmin_{1\leq s\leq S^*-1}\frac{\lambda_{s+1}}{\lambda_{s}},
\end{equation}
where $\lambda_1\geq\cdots\geq\lambda_{S^*}$ are the non-negative eigenvalues of $R$, which are obtained with matrix decomposition:
\begin{equation}
    R=V\diag\left(\lambda_1,\dots,\lambda_{S^*}\right)V^{-1},
    \label{eq:matrix_decomposition}
\end{equation}
where each row of $V\in\mathbb{R}^{S^*\times S^*}$ is the eigenvector that corresponds to the eigenvalues.
Note that the eigenvalues indicate the number of elements of each cluster where local attractors are softly assigned.

We used the hinge function to calculate the pairwise loss in \autoref{eq:pairwise_loss}, and we also know that attractors from the same block correspond to different speakers.
Thus, instead of $R$, we use the affinity matrix $R'=\left(r_{ij}'\right)$ defined as
\begin{align}
    r'_{ij}=\begin{cases}
    \mathbbm{1}\left(i=j\right)&\left(\text{$\vect{b}_i$ and $\vect{b}_j$ are from}\right.\\
    &\left.\text{the same block}\right)\\
    \frac{1}{1-\delta}\left[\operatorname{sim}\left(\vect{b}_i,\vect{b}_j\right)-\delta\right]_{+}&(\text{otherwise})
    \end{cases},\label{eq:modified_affinity_matrix}
\end{align}
where $\mathbbm{1}\left(\mathrm{cond}\right)$ is the indicator function that returns $1$ if $\mathrm{cond}$ is true and $0$ otherwise.
Matrix decomposition is then applied to $R'$ to obtain eigenvalues $\lambda'_1\geq\dots\geq\lambda'_{S^*}$ in the same manner as in \autoref{eq:matrix_decomposition}.
Although $R'$ is no longer positive-semidefinite, its eigenvalues are still good indicators of cluster size.
We only use the eigenvalues greater than or equal to one to estimate the number of speakers $\hat{S}$ as follows:
\begin{equation}
    \hat{S}=\argmin_{\substack{1\leq s\leq S^*-1 \\ \lambda'_s\geq1}}\frac{\lambda'_{s+1}}{\lambda'_{s}}.
\end{equation}

Although we set the affinity value between a pair of local attractors from the same block to be zero in \autoref{eq:modified_affinity_matrix}, naive clustering methods cannot force them to be assigned to different clusters.
Thus, we utilize a clustering method that can use cannot-link constraints.
COP-Kmeans clustering \cite{wagstaff2001constrained}, which is used in EEND-vector clustering \cite{kinoshita2021integrating,kinoshita2021advances} is one possible choice, but it sometimes results in failure because it cannot find the solution that fulfills the given constraints.
Thus, we use the CLC-Kmeans algorithm \cite{yang2013improved}, which is the modified version of the COP-Kmeans clustering, for inference of EEND-GLA.
To avoid having no solution due to cannot-link constraints, we update the estimated number of speakers before applying clustering as 
\begin{equation}
    \hat{S}\leftarrow\max\left(\hat{S},\max_{1\leq l\leq L} \hat{S}_l\right).
\end{equation}

EEND-GLA is optimized using both global- and local-attractor-based losses as in \autoref{eq:both_loss}, and we can use not only local attractors but also global attractors for inference.
Although local-attractor-based inference can deal with an arbitrary number of speakers, we found that global-attractor-based inference performs better when the number of speakers is low because it is trained in a fully supervised manner.
Therefore, we use the results from global and local attractors depending on the estimated number of speakers.
Assume that EEND-GLA is trained on mixtures each of which contains at most $N$ speakers.
If the estimated number of speakers using global attractors is less than $N$, we use the inference results based on global attractors. 
If it is equal to or larger than $N$, we use the inference results based on local attractors.
In this paper, the value of $N$ is set to four based on the simulated datasets we used for training, which are detailed in \autoref{sec:experimental_settings}. Even after the domain adaptation with real datasets with a larger number of speakers, we keep the value of $N$ unchanged during inference.

\subsection{Block-wise speaker-tracing buffer}\label{sec:bw_stb}
\begin{figure*}[t]
    \centering
    \includegraphics[width=1.0\linewidth]{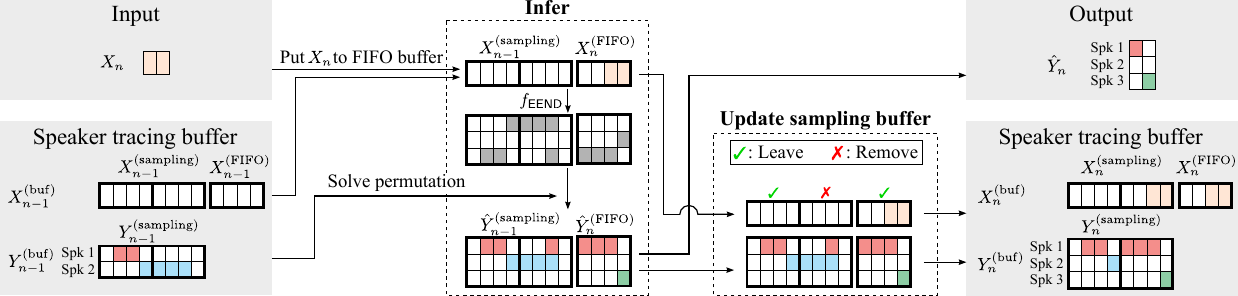}
    \caption{Online diarization using speaker-tracing buffer with block-wise update.}
    \label{fig:blockwise_stb}
\end{figure*}
As introduced in \autoref{sec:speaker_tracing_buffer}, the original speaker-tracing buffer (STB) includes a frame-wise (FW) selection step to meet the requirement of the buffer length.
Hereafter, for sake of distinction, we refer to it as FW-STB.
When trying to use FW-STB with EEND-GLA to perform online diarization of an unlimited number of speakers, the frame-wise selection can become a problem if the selected frames are not consecutive in the whole buffer.
The FIFO strategy ensures that the frames in the buffer are consecutive, but as mentioned in \autoref{sec:speaker_tracing_buffer}, it has difficulty in capturing long context.
On the other hand, while the sampling strategy can maintain long-range speaker consistency, the buffer can potentially contain non-consecutive frames of many different speakers; thus, the assumption of a limited number of speakers in a limited sequence of frames in the buffer does not hold.

To overcome this dilemma, we propose a block-wise speaker-tracing buffer (BW-STB).
The core idea of BW-STB is that it guarantees that the buffer consists of blocks, and each block contains the features and the corresponding results of consecutive frames.
If each block in the buffer is short enough that we can assume a limited amount of speakers, EEND-GLA can be used in the same way as the offline inference in which local attractors are obtained from the blocks formed by consecutive frames.
However, a naive implementation that waits for block-length features to accumulate and then processes them would result in block-length latency.
Thus, we use a frame-wise FIFO buffer and block-wise sampling buffer together to enable a low-latency online inference of EEND-GLA.

\autoref{fig:blockwise_stb} shows the proposed BW-STB. For simplicity, we assume that the buffer length $M$ is divisible by and longer than the block length $\lambda$, and $\lambda$ is divisible by the online processing unit $\nu$.
$M$-length BW-STB is divided into blocks of length $\lambda$ each. The first $\frac{M}{\lambda}-1$ is updated via block-wise sampling, and the last one is updated in a frame-wise FIFO manner.
We call them the sampling buffer and the FIFO buffer, respectively.
The features in BW-STB before the $n$-th input $X_n\in\mathbb{R}^{F\times\nu}$ can be written as 
\begin{align}
    X_{n-1}^\text{(buf)}&=\begin{bmatrix}X_{n-1}^\text{(samp)}&X_{n-1}^\text{(FIFO)}\end{bmatrix}\in\mathbb{R}^{F\times M},\\
    X_{n-1}^\text{(samp)}&=\begin{bmatrix}X_{n-1}^\text{(samp)}\left[1\right]\dots X_{n-1}^\text{(samp)}\left[\frac{M}{\lambda}-1\right]\end{bmatrix}\in\mathbb{R}^{F\times\left(M-\lambda\right)}.
\end{align}
Here, $X_{n-1}^\text{(samp)}$ are the features in the sampling buffer, where each $X_{n-1}^\text{(samp)}\left[k\right]\in\mathbb{R}^{F\times \lambda}~\left(k\in\left\{1,\dots,\frac{M}{\lambda}-1\right\}\right)$ are the features of consecutive $\lambda$ frames.
$X_{n-1}^\text{(FIFO)}\in\mathbb{R}^{F\times\lambda}$ is the features in the FIFO buffer, which contains those of the latest consecutive $\lambda$ frames.
In addition, each buffer contains the corresponding diarization results $Y_{n-1}^\text{(samp)}\in\left(0,1\right)^{\hat{S}_{n-1}\times \left(M-\lambda\right)}$ and $Y_{n-1}^\text{(FIFO)}\in\left(0,1\right)^{\hat{S}_{n-1}\times \lambda}$.

Given the input $X_n$, the features in the FIFO buffer are first updated as 
\begin{equation}
    X_{n}^\text{(FIFO)}=X_{n-1}^\text{(FIFO)}\begin{bmatrix}
    O_{\nu,\lambda-\nu}&O_{\nu,\nu}\\
    I_{\lambda-\nu}&O_{\lambda-\nu,\nu}%
    \end{bmatrix}+X_{n}\begin{bmatrix}
    O_{\nu,\lambda-\nu}&I_\nu%
    \end{bmatrix},
\end{equation}
where $I_a$ is an $a\times a$ identity matrix.
Note that the first $\lambda-\nu$ columns of $X_{n}^\text{(FIFO)}$ are identical to the last $\lambda-\nu$ columns of $X_{n-1}^\text{(FIFO)}$, and the last $\nu$ columns of $X_{n}^\text{(FIFO)}$ are identical to $X_{n}$.
Then, the diarization results are calculated from the concatenation of the features in the sampling and FIFO buffers as
\begin{equation}
    \begin{bmatrix}
    \hat{Y}_{n-1}^\text{(samp)}&\hat{Y}_{n}^\text{(FIFO)}%
    \end{bmatrix}
    =f_\mathsf{EEND}\left(\begin{bmatrix}
    X_{n-1}^\text{(samp)}&X_{n}^\text{(FIFO)}%
    \end{bmatrix}\right).
\end{equation}
With this estimation, the number of speakers is aligned via zero padding as described in \autoref{sec:speaker_tracing_buffer}, and then the speaker order of $\hat{Y}_{n-1}^\text{(samp)}$ and $\hat{Y}_{n}^\text{(FIFO)}$ is aligned to that of $Y_{n-1}^\text{(samp)}$ using \autoref{eq:solve_permutation}--\autoref{eq:frobenius_ip}.
Next, we output the last $\nu$ columns of the updated $Y_{n}^\text{(FIFO)}$, which correspond to the input $X_n$.

The sampling buffer is updated every time the FIFO buffer is fully replaced, \ie, after processing the $n$-th input where $n\equiv0\mod\frac{\lambda}{\nu}$.
During updates, $\frac{M}{\lambda}-1$ blocks are selected from $X_{n-1}^\text{(samp)}\left[1\right]\dots X_{n-1}^\text{(samp)}\left[\frac{M}{\lambda}-1\right]$ and $X_n^\text{(FIFO)}$, and they are stored as $X_{n}^\text{(samp)}$ in the sampling buffer.
The sampling probability of each block is calculated as a sum of $\tilde{\omega}_t$ of the frames in the block calculated using \autoref{eq:conventional_score_norm}.

With the aforementioned BW-STB, the online inference having the algorithmic latency of $\nu\left(\ll\lambda\right)$ is enabled.
Note that online diarization is performed using the FIFO buffer in the same way as FW-STB from the first to $\frac{\lambda}{\nu}$-th iterations because the sampling buffer is empty.

\subsection{Speaker-balanced sampling probabilities}
\begin{table}[t]
    \centering
    \caption{Example of sampling weights determined by (\ref{eq:conventional_score}) and (\ref{eq:proposed_score}).}
    \label{tbl:score_example}
    \resizebox{\linewidth}{!}{%
    \begin{tabular}{@{}ccccccccc@{}}
        \toprule
        &$t=1$&$t=2$&$t=3$&$t=4$&$t=5$&$t=6$&$t=7$&$t=8$\\\midrule
        $y_{1,t}$ & 0.999 & 0.999 & 0.999 & 0.999 & 0.999 & 0.001 & 0.001 & 0.001 \\
        $y_{2,t}$ & 0.001 & 0.001 & 0.001 & 0.001 & 0.001 & 0.001 & 0.001 & 0.999 \\\midrule
        $\tilde{\omega}_t$ by (\ref{eq:conventional_score_norm})(\ref{eq:conventional_score})&0.167&0.167&0.167&0.167&0.167&0.000&0.000&0.167\\
        $\tilde{\omega}_t$ by (\ref{eq:conventional_score_norm})(\ref{eq:proposed_score})&0.101&0.101&0.101&0.101&0.101&0.000&0.000&0.497\\
        \bottomrule
    \end{tabular}%
    }
\end{table}
The score in \autoref{eq:conventional_score} is designed to weigh more on frames where a single speaker dominates the conversation; as a result, the speaker-tracing buffer becomes informative enough to solve the speaker permutation ambiguity in \autoref{eq:solve_permutation}--\autoref{eq:permutation}.
However, in the case where some speakers dominate the conversation, the buffer contents might be biased toward those speakers, and hence the permutation ambiguity cannot be solved correctly.
For example, in the two-speaker example shown in \autoref{tbl:score_example}, $\tilde{\omega}_t$ is maximized at $t\in\left\{1,2,3,4,5,8\right\}$, where $\left(y_{1,t},y_{2,t}\right)\in\left\{\left(0.001,0.999\right),\left(0.999,0.001\right)\right\}$.
If $t=8$ is not selected to be stored in the buffer and the third speaker emerges in the next input, we cannot distinguish between the second and third speakers.

To make the buffer unbiased, we introduce the weighting factor $r_t$ into the sampling probability $\omega_t$ to balance the number of frames to be stored for each speaker.
We propose the following alternative:
\begin{equation}
    \omega_t=r_t\underbrace{\sum_{s=1}^{S_n}\bar{y}_{s,t}\log \left(\bar{y}_{s,t}S_n\right)}_{\autoref{eq:conventional_score}},\label{eq:proposed_score}
\end{equation}
where $r_t$ is defined as
\begin{equation}
    r_t=\sum_{s=1}^{S_n}\frac{y_{s,t}}{\sum_{t'=1}^T y_{s,t'}}.
\end{equation}
By this modification, in \autoref{tbl:score_example}, the sampling probability of $t=8$ becomes about a five times larger value (0.497) than that of $t\in\left\{1,2,3,4\right\}$ (0.101); thus, it is more likely to prevent the buffer from storing information that is biased toward the dominant speaker, \ie, the first speaker.

\subsection{Variable chunk-size training via minibatch reshaping}\label{sec:vct}
\begin{figure}[t]
    \centering
    \begin{minipage}{0.60\linewidth}
    \centering
    \subfloat[][Original minibatch]{\includegraphics[scale=0.8]{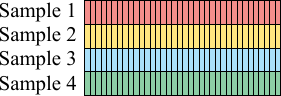}\label{fig:original_minibatch}}\\
    \subfloat[][VCT by masking]{\includegraphics[scale=0.8]{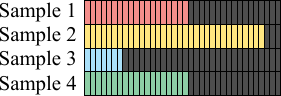}\label{fig:mask_vct}}
    \end{minipage}\hfill
    \begin{minipage}{0.40\linewidth}
    \centering
    \subfloat[][VCT by reshaping]{\includegraphics[scale=0.8]{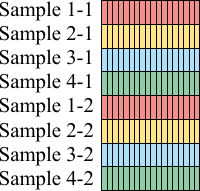}\label{fig:reshape_vct}}
    \end{minipage}
    \caption{Batch creation in the VCT.}
    \label{fig:vct}
\end{figure}

The VCT described in the last paragraph of \autoref{sec:speaker_tracing_buffer} varied the length of sequences by masking a part of each sequence (\autoref{fig:mask_vct}).
However, its calculation efficiency is low because the masked part does not contribute to the network optimization while still consuming GPU memory during training.

Therefore, we consider a method to use inputs of various lengths in the training process by reshaping the minibatch instead of masking.
If the minibatch at an iteration has minibatch size $B$ and input length $T$, we first reshape it to be a new minibatch with the size $B'=\frac{BT}{T'}$ and length $T'$, and then use it for training.
For \autoref{fig:vct}, the original minibatch has the size of four (\autoref{fig:original_minibatch}), and the reshaped minibatch has the size of eight by setting $T'=\frac{T}{2}$ (\autoref{fig:reshape_vct}).
In this paper, we set $B=64$ and $T=2000$, and in each iteration, with a probability of \SI{50}{\percent}, we set $T'$ to one of $\{50,100,200,500,1000\}$ to conduct VCT.
Let T' be a randomly selected value from {50,100,200,1000}.

\section{Experimental settings}\label{sec:experimental_settings}
\begin{table}[t]
    \centering
    \caption{Datasets used in our experiments.}
    \label{tbl:dataset}
    \captionsetup{position=top}
    \subfloat[][Simulated datasets.]{\label{tbl:simulated_dataset}%
    \begin{tabular}{@{}llccrr@{}}
        \toprule
        &&&&\multicolumn{1}{c}{Average}&\multicolumn{1}{c@{}}{Overlap}\\
        Dataset && \#Spk & \#Mixtures &\multicolumn{1}{c}{duration}& \multicolumn{1}{c@{}}{ratio}\\\midrule
        \textbf{Train}&Sim1spk& 1 & 100,000&\SI{76.8}{\second}&\SI{0.0}{\percent}\\
        &Sim2spk& 2 & 100,000&\SI{88.6}{\second}&\SI{34.1}{\percent}\\
        &Sim3spk& 3 & 100,000&\SI{151.2}{\second}&\SI{34.2}{\percent}\\
        &Sim4spk& 4 & 100,000&\SI{238.1}{\second}&\SI{31.5}{\percent}\\\midrule
        \textbf{Adaptation}&Sim1spk& 1 & 1,000&\SI{76.0}{\second}&\SI{0.0}{\percent}\\
        &Sim2spk& 2 & 1,000&\SI{89.3}{\second}&\SI{34.5}{\percent}\\
        &Sim3spk& 3 & 1,000&\SI{150.3}{\second}&\SI{34.9}{\percent}\\
        &Sim4spk& 4 & 1,000&\SI{238.2}{\second}&\SI{31.4}{\percent}\\\midrule
        \textbf{Test}&Sim1spk& 1 & 500&\SI{77.2}{\second}&\SI{0.0}{\percent}\\
        &Sim2spk& 2 & 500&\SI{88.2}{\second}&\SI{34.4}{\percent}\\
        &Sim3spk& 3 & 500&\SI{149.7}{\second}&\SI{34.7}{\percent}\\
        &Sim4spk& 4 & 500&\SI{237.4}{\second}&\SI{32.0}{\percent}\\
        &Sim5spk& 5 & 500&\SI{328.8}{\second}&\SI{30.7}{\percent}\\
        &Sim6spk& 6 & 500&\SI{423.4}{\second}&\SI{29.9}{\percent}\\
        \bottomrule
    \end{tabular}
    }\\
    \subfloat[][Real datasets.]{\label{tbl:real_dataset}%
    \resizebox{\linewidth}{!}{%
    \begin{tabular}{@{}llccccc@{}}
        \toprule
        &&&&&\multicolumn{1}{c}{Average}&\multicolumn{1}{c@{}}{Overlap}\\
        Dataset && Split&\#Spk & \#Mixtures & \multicolumn{1}{c}{duration}& \multicolumn{1}{c@{}}{ratio}\\\midrule
        \textbf{Adaptation}&CALLHOME \cite{callhome}&Part 1&2--7&249&\SI{125.8}{\second}&\SI{17.0}{\percent}\\
        &DIHARD II \cite{ryant2019second}&dev&1--10&192&\SI{444.8}{\second}&\SI{9.8}{\percent}\\
        &DIHARD III \cite{ryant2021third}&dev&1--10&254&\SI{483.4}{\second}&\SI{10.7}{\percent}\\\midrule
        \textbf{Test}&CALLHOME \cite{callhome} &Part 2&2--6 & 250 & \SI{123.2}{\second}&\SI{16.7}{\percent}\\
        &DIHARD II \cite{ryant2019second}&eval &1--9 &194&\SI{416.3}{\second}&\SI{8.9}{\percent}\\
        & DIHARD III \cite{ryant2021third}&eval&1--9&259&\SI{458.1}{\second}&\SI{9.2}{\percent}\\
        \bottomrule
    \end{tabular}%
    }
    }
\end{table}

The initial training of each EEND-based model was based on the simulated mixtures shown in \autoref{tbl:simulated_dataset}.
These were made with NIST SRE and Switchboard datasets as speech corpora, MUSAN \cite{snyder2015musan} as a noise corpus, and simulated room impulse responses \cite{ko2017study}, following the protocol used for the original EEND \cite{fujita2019end2}.
Following our previous studies \cite{horiguchi2022encoderdecoder,horiguchi2021towards}, we first trained each EEND-based model using Sim2spk from scratch for 100 epochs and then finetuned it using the concatenation of Sim\{1,2,3,4\}spk for another 50 epochs.
The Adam optimizer \cite{kingma2015adam} with Noam scheduler \cite{vaswani2017attention} was used during the training using the simulated datasets.
For online purposes, the model was adapted using the adaptation set of Sim\{1,2,3,4\}spk for an additional 100 epochs using variable chunk-size training (VCT).
This time, the Adam optimizer with a fixed learning rate of $1\times10^{-5}$ was used.
Note that the adaptation set of each simulated dataset was the subset of the corresponding training set.
The training process took about two weeks with a single NVIDIA\textsuperscript{\textregistered} Tesla\textsuperscript{\textregistered} V100 GPU.

We also used the real datasets shown in \autoref{tbl:real_dataset} for evaluation.
The model pretrained using Sim\{1,2,3,4\}spk was further adapted to the CALLHOME, DIHARD II, and DIHARD III datasets, respectively.
The adaptation was conducted for another 100 epochs using the Adam optimizer with a learning rate of $1\times10^{-5}$.
For online purposes, VCT was used instead.

For EEND-GLA, we used four- or six-stacked Transformer encoders, each outputting 256-dimensional frame-wise embeddings.
We call each EEND-GLA-Small and EEND-GLA-Large, respectively.
For EDA, we used an encoder-decoder based on single-layer long short-term memory with 256-dimensional hidden units. Note that the order of the input sequence is shuffled before being fed into EDA following the conventional study \cite{horiguchi2022encoderdecoder}.
For the inputs to the models, 345-dimensional acoustic features extracted for each \SI{100}{\ms} were used, and they were obtained in the following steps: 1) extract 23-dimensional log-mel filterbanks for every \SI{10}{\ms}, 2) apply frame splicing ($\pm 7$ frames), and 3) subsample by a factor of 10.

Unless otherwise specified, the length of an online processing unit $\nu$ was set to \SI{1}{\second}, and the buffer length was set to \SI{100}{\second}.
The block length $\lambda$ of the BW-STB was set to \SI{5}{\second} following EEND-GLA \cite{horiguchi2021towards}; as a result, the length of sampling and FIFO buffers are \SI{95}{\second} and \SI{5}{\second}, respectively.

For evaluating offline diarization, we utilized several cascaded methods \cite{landini2022bayesian,landini2020but,bredin2021endtoend,horiguchi2021hitachi,cheng2022multi} and end-to-end methods \cite{horiguchi2020endtoend,horiguchi2022encoderdecoder,fujita2020neural,kinoshita2021advances} for comparison.
For evaluating online diarization, we used FW-STB with EEND-EDA based on four-stacked Transformers \cite{xue2021online2}.
In addition, we referred to the results of various conventional online diarization methods \cite{zhang2019fully,fini2020supervised,zhang2022online,xue2021online2,han2021bwedaeend,coria2021overlapaware,yue2022online} on various datasets.
Some cascaded comparison methods \cite{zhang2019fully,fini2020supervised,zhang2022online} used the oracle SAD; for a fair comparison, we used SAD post-processing \cite{horiguchi2022encoderdecoder} for the results of EEND-based methods to recover missed speech and filter false-alarmed speech.

For the evaluation protocol, we used DERs.
Following the previous studies \cite{horiguchi2022encoderdecoder}, we forgave \SI{0.25}{\second} of its collar tolerance in the evaluations of the simulated datasets and CALLHOME, while we did not allow such a collar in the evaluations of the DIHARD II and DIHARD III datasets.

\section{Results}\label{sec:results}
\subsection{Evaluation of the variations of speaker-tracing buffer}\label{sec:eval_stb}
Before we dive into the evaluation of EEND-GLA, we evaluated the effects of each modification on the speaker-tracing buffer using EEND-EDA.
Step-by-step improvement on the CALLHOME dataset is shown in \autoref{tbl:result_eend_eda_callhome}.
The DERs were significantly reduced by using VCT.
In a comparison of the results in the third and fourth lines, VCT by reshaping outperformed that by masking in all the conditions.
Introducing the speaker-balancing term in the sampling probability as in \autoref{eq:proposed_score} improved the DERs except when the buffer length was too short to store enough information to solve the speaker permutation ambiguity, as in the fifth line.
Finally, replacing FW-STB with BW-STB did not affect the diarization performance as shown in the last line, except when the buffer length was \SI{10}{\second}, where the sampling buffer consisted of only one block.

\begin{table}[t]
    \centering
    \caption{Step-by-step improvement in the online inference of EEND-EDA on the CALLHOME dataset. VCT: Variable chunk-size training.}
    \setlength{\tabcolsep}{3pt}
    \label{tbl:result_eend_eda_callhome}
    \begin{threeparttable}
    \resizebox{\linewidth}{!}{%
    \begin{tabular}{@{}lcccccccccc@{}}
        \toprule
        &&&\multicolumn{8}{c}{Buffer length (\si{\second})}\\\cmidrule(l{3pt}){4-11}
        &VCT&$\omega_t$&1&2&5&10&20&50&100&$\infty$\\\midrule
        FW-STB \cite{xue2021online2} \tnote{\dag}&None&(\ref{eq:conventional_score})&N/A&N/A&N/A&26.6&N/A&20.0&19.5&N/A\\
        FW-STB \cite{xue2021online2}&None&(\ref{eq:conventional_score})&89.79&76.98&43.87&28.13&21.82&19.69&18.54&18.34\\
        FW-STB &Mask&(\ref{eq:conventional_score})&50.56&44.95&27.48&21.53&18.12&16.82&16.78&15.69\\
        FW-STB &Reshape&(\ref{eq:conventional_score})&\textbf{44.11}&\textbf{36.61}&25.41&20.54&18.11&16.20&15.74&\textbf{15.00}\\
        FW-STB &Reshape&(\ref{eq:proposed_score})&45.60&37.36&\textbf{24.19}&\textbf{20.10}&\textbf{16.79}&15.50&\textbf{14.93}&\textbf{15.00}\\
        BW-STB &Reshape&(\ref{eq:proposed_score})&N/A&N/A&N/A&24.27&16.84&\textbf{15.03}&15.06&15.42\\
        \bottomrule
    \end{tabular}%
    }
    \begin{tablenotes}
        \item[\dag] The values are from the original FW-STB paper \cite{xue2021online2}.
    \end{tablenotes}
    \end{threeparttable}
\end{table}

\begin{figure}[t]
    \centering
    \subfloat[][DER]{\includegraphics[width=0.48\linewidth]{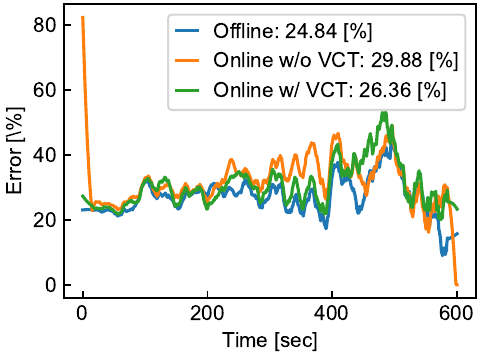}}\hfill
    \subfloat[][Miss]{\includegraphics[width=0.48\linewidth]{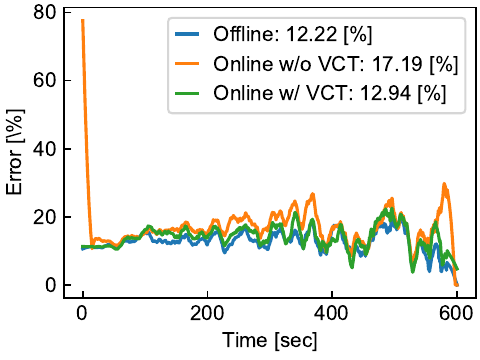}}\\
    \subfloat[][False alarm]{\includegraphics[width=0.48\linewidth]{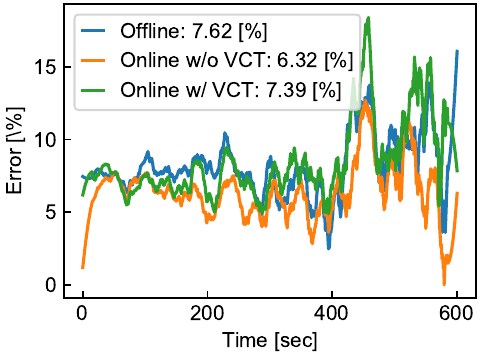}}\hfill
    \subfloat[][Speaker confusion]{\includegraphics[width=0.48\linewidth]{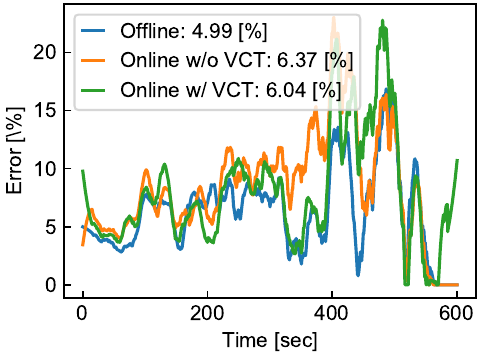}}
    \caption{Frame-wise breakdown of diarization error on CALLHOME.}
    \label{fig:error_analysis_callhome}
\end{figure}

For the detailed error analyses, we show the frame-level breakdown of the diarization error of FW-STB with and without VCT in \autoref{fig:error_analysis_callhome}.
Each graph was smoothed along the time axis using the Savizky-Golay filter \cite{savitzky1964smoothing} for visualization purposes.
We clearly observed that VCT drastically decreased the error caused by missed speech at the very beginning of recordings with a slight increase in false alarms.
Note that the DERs shown in \autoref{fig:error_analysis_callhome} are different from those in \autoref{tbl:result_eend_eda_callhome} because the results are without a collar.

In the following experiments, we used FW-STB and BW-STB in the last two lines in \autoref{tbl:result_eend_eda_callhome}, \ie, VCT by reshaping and speaker-balanced sampling probabilities were utilized.

\subsection{Evaluation of offline and online diarization for an unlimited number of speakers}
\subsubsection{Simulated dataset}
We first evaluated EEND-GLA on the simulated datasets. The results are shown in \autoref{tbl:result_simu}.

For the evaluation of offline processing, Kaldi's x-vector clustering recipe\footnote{\url{https://github.com/kaldi-asr/kaldi/tree/master/egs/callhome_diarization/v2}} was used as a baseline. The x-vector extractor was trained using the same set of datasets that were used to create the simulated datasets in \autoref{tbl:simulated_dataset}.
Note that the baseline has no way to handle overlapping speech.
EEND-EDA and EEND-GLA-Small performed evenly on the datasets of the seen number of speakers, while EEND-GLA-Small significantly outperformed EEND-EDA on the datasets of the unseen number of speakers.
It clearly showed that EEND-GLA-Small could deal with a higher number of speakers than that observed during training by introducing clustering.
It is worth mentioning that EEND-EDA sometimes outputs more than four attractors, but the results in \autoref{tbl:result_simu} in which ignoring the fifth and subsequent attractors improved the DERs indicate that these attractors were not correctly calculated to represent the fifth and subsequent speakers.
Using EEND-GLA-Large improved the DERs for the seen number of speakers, but those for the unseen number of speakers were degraded.
We considered this to be because the network was overtrained on the seen number of speakers with the larger model.
For comparison, we also showed the DERs of EEND-EDA trained using mixtures, each of which contained at most five speakers.
It showed a better DER on five-speaker mixtures, but the DER on six-speaker mixtures degraded rapidly.
EEND-GLA achieved DERs comparable to EEND-EDA for five-speaker mixtures and significantly outperformed it for six-speaker mixtures.

In terms of online processing, STB-based methods outperformed BW-EDA-EEND \cite{han2021bwedaeend} in all but single-speaker data even though the online processing unit was \SI{1}{\second}, which was ten times shorter than that of BW-EEND-EDA.
Online inference of EEND-GLA-Small using BW-STB significantly improved DERs on five- and six-speaker mixtures, which were not observed during training.
EEND-GLA-Large improved the DERs for the seen number of speaker conditions of EEND-GLA-Small but degraded the DERs for the unseen number of speaker conditions, the same as in offline inference.

\begin{table}[t]
    \centering
    \caption{DERs (\%) on the simulated datasets with \SI{0.25}{\second} collar tolerance. Unless otherwise specified, each online system had an algorithmic latency of \SI{1}{\second}.}
    \setlength{\tabcolsep}{3pt}
    \label{tbl:result_simu}
    \resizebox{\linewidth}{!}{%
    \begin{threeparttable}
    \begin{tabular}{@{}lcccccc@{}}
        \toprule
        &\multicolumn{6}{c}{\# of speakers}\\\cmidrule(l{3pt}){2-7}
        &\multicolumn{4}{c}{seen}&\multicolumn{2}{c}{unseen}\\\cmidrule(l{3pt}r{3pt}){2-5}\cmidrule(l{3pt}){6-7}
        &1&2&3&4&5&6\\\midrule
        \textbf{Offline}\\
        \quad X-vector clustering&37.42&7.74&11.46&22.45&31.00&38.62\\
        \quad EEND-EDA \cite{horiguchi2020endtoend,horiguchi2022encoderdecoder} &0.15&\textbf{3.19}&6.60&9.26&23.11&34.97\\
        \quad EEND-EDA \cite{horiguchi2020endtoend,horiguchi2022encoderdecoder} \tnote{\dag} &0.15&\textbf{3.19}&6.60&8.68&22.43&33.28\\
        \quad EEND-GLA-Small&0.25&3.53&6.79&8.98&\textbf{12.44}&\textbf{17.98}\\
        \quad EEND-GLA-Large&\textbf{0.09}&3.54&\textbf{5.74}&\textbf{6.79}&12.51&20.42\\\cmidrule(l{1em}){1-7}
        \quad EEND-EDA \cite{horiguchi2020endtoend,horiguchi2022encoderdecoder} \tnote{\ddag}&0.36&3.65&7.70&9.97&11.95&22.59\\
        \midrule
        \textbf{Online}\\
        \quad BW-EDA-EEND \cite{han2021bwedaeend}\tnote{\S}&\textbf{1.03}&6.10&12.58&19.17&N/A&N/A\\
        \quad EEND-EDA \cite{horiguchi2020endtoend,horiguchi2022encoderdecoder} + FW-STB&1.50&5.91&9.79&11.92&26.57&37.31\\
        \quad EEND-EDA \cite{horiguchi2020endtoend,horiguchi2022encoderdecoder} + FW-STB \tnote{\dag}&1.50&5.91&9.79&11.85&26.63&37.25\\
        \quad EEND-GLA-Small + BW-STB&1.19&5.18&9.41&13.19&\textbf{16.95}&\textbf{22.55}\\
        \quad EEND-GLA-Large + BW-STB&1.12&\textbf{4.61}&\textbf{8.14}&\textbf{11.38}&17.27&25.77\\\cmidrule(l{1em}){1-7}
        \quad EEND-EDA \cite{horiguchi2020endtoend,horiguchi2022encoderdecoder} + FW-STB \tnote{\ddag}&1.33&6.01&10.49&12.64&15.28&26.09\\
        \bottomrule
    \end{tabular}%
    \begin{tablenotes}
        \item[\dag] Four attractors were used at most.
        \item[\ddag] Trained on Sim\{1,2,3,4,5\}spk. Five attractors were used at most.
        \item[\S] Algorithmic latency \SI{10}{\second}.
    \end{tablenotes}
    \end{threeparttable}%
    }
\end{table}

\autoref{tbl:result_simu_loss} shows the offline DERs of EEND-GLA-Small obtained with various training and inference strategies.
Even when only local attractors were used during both training and inference, it achieved better DERs than EEND-EDA for the unseen numbers of speakers but worse ones for the seen numbers of speakers (first row).
Using the global attractors jointly for training improved the performance for the seen numbers of speakers, but it was still not as good as EEND-EDA when only the local attractors were used for inference (second row), especially when the number of speakers was low (\ie, one- or two-speaker cases).
This is because a small error in the number of speakers (\eg, $\pm 1$) led to a high degradation of DER.
Using the results based on global attractors when the number of speakers was low resulted in good DERs for both seen and unseen numbers of speakers (third row). 

\begin{table}[t]
    \centering
    \caption{Offline DERs (\%) of EEND-GLA-Small with various training and inference strategies. Loss: the training objective used for training. Inference: attractors used during inference.}
    \setlength{\tabcolsep}{3pt}
    \label{tbl:result_simu_loss}
    \resizebox{\linewidth}{!}{%
    \begin{tabular}{@{}llcccccc@{}}
        \toprule
        &&\multicolumn{6}{c}{\# of speakers}\\\cmidrule(l{3pt}){3-8}
        &&\multicolumn{4}{c}{Seen}&\multicolumn{2}{c}{Unseen}\\\cmidrule(l{3pt}r{3pt}){3-6}\cmidrule(l{3pt}){7-8}
        Loss&Inference&1&2&3&4&5&6\\\midrule
        $\mathcal{L}_\text{local}$ \autoref{eq:local_loss}&Local&8.85&12.71&10.31&11.14&14.11&19.36\\
        $\mathcal{L}_\text{local}+\mathcal{L}_\text{global}$ \autoref{eq:both_loss}&Local&2.84&10.21&7.54&9.08&\textbf{12.40}&18.03\\
        $\mathcal{L}_\text{local}+\mathcal{L}_\text{global}$ \autoref{eq:both_loss}&Local \& Global&\textbf{0.25}&\textbf{3.53}&\textbf{6.79}&\textbf{8.98}&12.44&\textbf{17.98}\\
        \bottomrule
    \end{tabular}%
    }
\end{table}

We also show the confusion matrices for speaker counting on the simulated datasets in \autoref{tbl:speaker_counting_simu}.
The speaker counting accuracy of EEND-GLA-Small with BW-STB outperformed that of EEND-EDA with FW-STB, and the gaps between them were larger especially when the number of speakers was higher than four.
Note that EEND-EDA with FW-STB sometimes produced the results of more than four speakers, but they did not help estimate the speech activities of more than four speakers correctly as we stated in this section.

\begin{table}[t]
    \centering
    \caption{Confusion matrices for speaker counting on the simulated datasets.}
    \label{tbl:speaker_counting_simu}
    \captionsetup{position=top}
    \setlength{\tabcolsep}{3pt}
    \subfloat[][EEND-EDA (offline)]{%
        \resizebox{0.48\linewidth}{!}{%
        \begin{tabular}{@{}cc|cccccc@{}}
            \toprule
            &&\multicolumn{6}{c}{Ref. \#Speakers}\\
            &&1&2&3&4&5&6\\\midrule
            \multirow{7}{*}{\rotatebox{90}{Pred. \#Speakers}}&1 & \textbf{500}&0&0&0&0&0\\
            &2 & 0&\textbf{482}&0&0&0&0\\
            &3 & 0&17&\textbf{435}&5&1&0\\
            &4 & 0&1&65&\textbf{447}&224&139\\
            &5 & 0&0&0&48&\textbf{268}&337\\
            &6 & 0&0&0&0&7&\textbf{24}\\
            &7+ & 0&0&0&0&0&0\\
            \bottomrule
        \end{tabular}%
    }%
    }
    \hfill
    \subfloat[][EEND-GLA-Small (offline)]{%
        \resizebox{0.48\linewidth}{!}{%
        \begin{tabular}{@{}cc|cccccc@{}}
            \toprule
            &&\multicolumn{6}{c}{Ref. \#Speakers}\\
            &&1&2&3&4&5&6\\\midrule
            \multirow{7}{*}{\rotatebox{90}{Pred. \#Speakers}}&1 & \textbf{498}&0&0&0&0&0\\
            &2 & 2&\textbf{474}&0&0&0&0\\
            &3 & 0&25&\textbf{451}&17&2&1\\
            &4 & 0&1&33&\textbf{412}&78&30\\
            &5 & 0&0&10&62&\textbf{361}&183\\
            &6 & 0&0&6&7&47&\textbf{229}\\
            &7+ & 0&0&0&2&12&57\\
            \bottomrule
        \end{tabular}%
    }%
    }\\
    \subfloat[][EEND-EDA + FW-STB (online)]{%
        \resizebox{0.48\linewidth}{!}{%
        \begin{tabular}{@{}cc|cccccc@{}}
            \toprule
            &&\multicolumn{6}{c}{Ref. \#Speakers}\\
            &&1&2&3&4&5&6\\\midrule
            \multirow{7}{*}{\rotatebox{90}{Pred. \#Speakers}}
            &1 & \textbf{376}&0&0&0&0&0\\
            &2 & 120&\textbf{244}&0&0&0&0\\
            &3 & 4&249&\textbf{252}&1&0&0\\
            &4 & 0&7&245&\textbf{449}&271&172\\
            &5 & 0&0&3&50&\textbf{222}&314\\
            &6 & 0&0&0&0&7&\textbf{14}\\
            &7+& 0&0&0&0&0&0\\
            \bottomrule
        \end{tabular}%
        }%
    }
    \hfill
    \subfloat[][EEND-GLA-Small + BW-STB (online)\label{tbl:speaker_counting_simu_bw}]{%
        \resizebox{0.48\linewidth}{!}{%
        \begin{tabular}{@{}cc|ccccccc@{}}
            \toprule
            &&\multicolumn{6}{c}{Ref. \#Speakers}\\
            &&1&2&3&4&5&6\\\midrule
            \multirow{7}{*}{\rotatebox{90}{Pred. \#Speakers}}
            &1 & \textbf{411}&0&0&0&0&0\\
            &2 & 84&\textbf{343}&0&0&0&0\\
            &3 & 5&156&\textbf{370}&3&0&0\\
            &4 & 0&1&109&\textbf{302}&16&0\\
            &5 & 0&0&20&181&\textbf{364}&38\\
            &6 & 0&0&1&13&114&\textbf{385}\\
            &7+& 0&0&0&1&6&77\\
            \bottomrule
        \end{tabular}%
        }%
    }
\end{table}

\subsubsection{CALLHOME}
\autoref{tbl:results_callhome} shows the DERs on the CALLHOME dataset.
In the evaluation of offline processing, EEND-GLA-Small and EEND-GLA-Large outperformed the conventional methods with \SI{11.92}{\percent} and \SI{11.84}{\percent} DERs, respectively.

In online diarization, compared with the original FW-STB \cite{xue2021online2}, our updates on VCT and the sampling probabilities improved the DERs from \SI{19.51}{\percent} to \SI{14.93}{\percent}.
EEND-GLA-Small and EEND-GLA-Large with BW-STB further improved DERs to \SI{14.80}{\percent} and \SI{14.29}{\percent}, respectively.
Our method also outperformed BW-EDA-EEND \cite{han2021bwedaeend} by a large margin.

\begin{table}[t]
    \centering
    \caption{DERs (\%) on CALLHOME with \SI{0.25}{\second} collar tolerance. Unless otherwise specified, each online system had an algorithmic latency of \SI{1}{\second}.}
    \setlength{\tabcolsep}{3pt}
    \label{tbl:results_callhome}
    \resizebox{\linewidth}{!}{%
    \begin{threeparttable}
    \begin{tabular}{@{}lcccccc@{}}
        \toprule
        &\multicolumn{5}{c}{\# of speakers}\\\cmidrule(l{3pt}r{3pt}){2-6}
        Method&2&3&4&5&6&All\\\midrule
        \textbf{Offline}\\
        \quad VBx \cite{landini2022bayesian} \tnote{\dag}& 9.44&13.89&16.05&\textbf{13.87}&24.73&13.28\\
        \quad MTFAD \cite{cheng2022multi}&N/A&N/A&N/A&N/A&N/A&14.31\\
        \quad SC-EEND \cite{fujita2020neural}&9.57&14.00&21.14&31.07&37.06&15.75\\
        \quad EEND-EDA \cite{horiguchi2020endtoend,horiguchi2022encoderdecoder}&7.83&12.29&17.59&27.66&37.17&13.65\\
        \quad EEND-vector clust. \cite{kinoshita2021advances} &7.94&11.93&16.38&21.21&23.10&12.49\\
        \quad EEND-VC-iGMM \cite{kinoshita2022tight}&8.6&12.6&16.1&27.5&26.9&13.3\\
        \quad EEND-GLA-Small&\textbf{6.94}&\textbf{11.42}&14.49&29.76&24.09&11.92\\
        \quad EEND-GLA-Large&7.11&11.88&\textbf{14.37}&25.95&\textbf{21.95}&\textbf{11.84}\\\midrule
        \textbf{Online}\\
        \quad BW-EDA-EEND \cite{han2021bwedaeend} \tnote{\ddag}& 11.82&18.30&25.93&N/A&N/A&N/A\\
        \quad EEND-EDA \cite{horiguchi2020endtoend,horiguchi2022encoderdecoder} + FW-STB \tnote{\S}&12.70&18.40&24.30&35.83&42.21&19.51\\
        \quad EEND-EDA \cite{horiguchi2020endtoend,horiguchi2022encoderdecoder} + FW-STB&9.08&13.33&19.36&30.09&37.21&14.93\\
        \quad EEND-GLA-Small + BW-STB &\textbf{9.01}&12.73&19.45&32.26&36.78&14.80\\
        \quad EEND-GLA-Large + BW-STB&9.20&\textbf{12.42}&\textbf{18.21}&\textbf{29.54}&\textbf{35.03}&\textbf{14.29}\\
        \bottomrule
    \end{tabular}%
    \begin{tablenotes}
        \item[\dag] The oracle SAD was used.
        \item[\ddag] Algorithmic latency of \SI{10}{\second}.
        \item[\S] The values are from the original FW-STB paper \cite{xue2021online2}.
    \end{tablenotes}
    \end{threeparttable}%
    }
\end{table}

\subsubsection{DIHARD II and III}
\begin{table}[t]
    \centering
    \caption{DERs (\%) on DIHARD II with no collar tolerance. Each online system had an algorithmic latency of \SI{1}{\second}.}
    \label{tbl:results_dihard2}
    \begin{threeparttable}
    \begin{tabular}{@{}lccc@{}}
        \toprule
        &\multicolumn{2}{c}{\# of speakers}\\\cmidrule(l{3pt}r{3pt}){2-3}
        Method&$\leq4$&$\geq5$&All\\\midrule
        \textbf{Offline}\\
        \quad BUT system \cite{landini2020but} & \textbf{21.34}&39.85&27.11\\
        \quad VBx + overlap-aware resegmentation \cite{bredin2021endtoend}&21.41&\textbf{36.93}&\textbf{26.25}\\
        \quad EEND-EDA \cite{horiguchi2020endtoend}&22.09&47.66&30.07\\
        \quad EEND-GLA-Small&22.24&44.92&29.31\\
        \quad EEND-GLA-Large&21.40&43.62&28.33\\
        \midrule
        \textbf{Online}\\
        \quad Overlap-aware speaker embeddings \cite{coria2021overlapaware}&27.00&52.62&34.99\\
        \quad EEND-EDA \cite{horiguchi2020endtoend,horiguchi2022encoderdecoder} + FW-STB \tnote{\dag}&28.14&53.64&36.09\\
        \quad EEND-EDA \cite{horiguchi2020endtoend,horiguchi2022encoderdecoder} + FW-STB&25.63&50.45&33.37\\
        \quad EEND-GLA-Small + BW-STB&23.96&48.06&31.47\\
        \quad EEND-GLA-Large + BW-STB&\textbf{22.62}&\textbf{47.06}&\textbf{30.24}\\
        \textbf{Online (with oracle voice activity detection)}\\
        \quad UIS-RNN \cite{zhang2019fully}&N/A&N/A&30.9\\
        \quad UIS-RNN-SML \cite{fini2020supervised}&N/A&N/A&27.3\\
        \quad Core samples selection \cite{yue2022online}&N/A&N/A&23.1\\
        \quad EEND-EDA \cite{horiguchi2020endtoend,horiguchi2022encoderdecoder} + FW-STB \tnote{\dag}&17.21&43.58&25.44\\
        \quad EEND-EDA \cite{horiguchi2020endtoend,horiguchi2022encoderdecoder} + FW-STB&16.56&42.58&24.67\\
        \quad EEND-GLA-Small + BW-STB&15.29&40.85&23.26\\
        \quad EEND-GLA-Large + BW-STB&\textbf{13.55}&\textbf{40.39}&\textbf{21.92}\\
        \bottomrule
    \end{tabular}%
    \begin{tablenotes}
        \item[\dag] The values are from the original FW-STB paper \cite{xue2021online2}.
    \end{tablenotes}
    \end{threeparttable}
\end{table}

\autoref{tbl:results_dihard2} shows the results on the DIHARD II dataset.
In offline diarization, EEND-GLA-Small and EEND-GLA-Large improved the DERs from EEND-EDA, especially when the number of speakers was higher than four.
Compared with the cascaded method \cite{landini2021but} or the cascaded method incorporated with EEND for post-processing \cite{bredin2021endtoend}, EEND-GLA-Large performed on par with them when the number of speakers was low, but not when the number of speakers was high. 

In online diarization, the DER of EEND-EDA was improved by using the proposed FW-STB from \SI{36.09}{\percent} to \SI{33.37}{\percent}, and BW-STB further improved the DERs to \SI{31.47}{\percent} and \SI{30.24}{\percent} with EEND-GLA-Small and EEND-GLA-Large, respectively.
If we focus on the comparison methods, overlap-aware speaker embedding \cite{bredin2021endtoend,coria2021overlapaware} had a large gap in the DERs between offline and online inference (\SI{26.25}{\percent} vs. \SI{34.99}{\percent}).
This is because its offline performance was highly boosted by using VBx \cite{landini2022bayesian}, which is not suited for online inference.
However, the gap between the DERs of offline and online inference of EEND-GLA was only about two points and outperformed the comparable method in both cases where the number of speakers was low or high.
We also show the DERs with UIS-RNN \cite{zhang2019fully} and UIS-RNN-SML \cite{fini2020supervised}, which are based on fully supervised clustering of d-vectors extracted using a sliding window, under the condition that the oracle SAD was used.
In this case, too, the EEND-GLA-based methods outperformed these comparable methods.

\begin{table}[t]
    \centering
    \caption{DERs (\%) on DIHARD III with no collar tolerance. Unless otherwise specified, each online system had an algorithmic latency of \SI{1}{\second}.}
    \label{tbl:results_dihard3}
    \begin{threeparttable}
    \begin{tabular}{@{}lccc@{}}
        \toprule
        &\multicolumn{2}{c}{\# of speakers}\\\cmidrule(l{3pt}r{3pt}){2-3}
        Method&$\leq4$&$\geq5$&All\\\midrule
        \textbf{Offline}\\
        \quad VBx + overlap handling \cite{horiguchi2021hitachi} & 16.38&42.51& 21.47\\
        \quad VBx + overlap-aware resegmentation \cite{bredin2021endtoend} &15.32&\textbf{35.87}&\textbf{19.33}\\
        \quad EEND-EDA \cite{horiguchi2020endtoend,horiguchi2022encoderdecoder} &15.55&48.30  & 21.94\\
        \quad EEND-GLA-Small&14.39& 44.32& 20.23\\
        \quad EEND-GLA-Large&\textbf{13.64}&43.67&19.49\\\midrule
        \textbf{Online}\\
        \quad Overlap-aware speaker embeddings \cite{coria2021overlapaware}&21.07&54.28&27.55\\
        \quad EEND-EDA \cite{horiguchi2020endtoend,horiguchi2022encoderdecoder} + FW-STB&19.00&50.21&25.09\\
        \quad EEND-GLA-Small + BW-STB&15.87&47.27&22.00\\
        \quad EEND-GLA-Large + BW-STB&\textbf{14.81}&\textbf{45.17}&\textbf{20.73}\\
        \textbf{Online (with oracle voice activity detection)}\\
        \quad System by Zhang \etal~\cite{zhang2022online}~\tnote{\dag}&N/A&N/A&19.57\\
        \quad Core samples selection \cite{yue2022online}&N/A&N/A&19.3\\
        \quad EEND-EDA \cite{horiguchi2020endtoend,horiguchi2022encoderdecoder} + FW-STB& 12.80&42.46&18.58\\
        \quad EEND-GLA-Small + BW-STB& 9.91&40.21&15.82\\
        \quad EEND-GLA-Large + BW-STB& \textbf{8.85}&\textbf{38.86}&\textbf{14.70}\\
        \bottomrule
    \end{tabular}
    \begin{tablenotes}
        \item[\dag] Algorithmic latency of \SI{0.5}{\second}.
    \end{tablenotes}
    \end{threeparttable}%
\end{table}

We also show the DERs on the DIHARD III dataset in \autoref{tbl:results_dihard3}.
The results were almost the same as those of the DIHARD II dataset.
EEND-GLA-Large achieved \SI{19.49}{\percent} DER in offline diarization, which was as accurate as the best performing conventional method \cite{bredin2021endtoend}, and \SI{20.73}{\percent} DER in online diarization, which was about seven points better than that of the conventional method \cite{coria2021overlapaware}.

\subsection{Real time factor}
\begin{figure}
    \centering
    \includegraphics[width=0.85\linewidth]{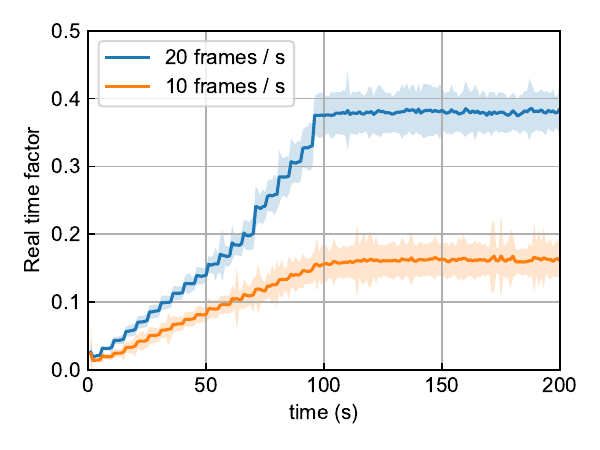}
    \caption{Real time factor of EEND-GLA-Small with BW-STB calculated using Sim5spk. The filled areas represent the standard deviations. The DERs are \SI{16.95}{\percent} and \SI{18.18}{\percent} with 10 frames/s and 20 frames/s conditions, respectively.}
    \label{fig:rtf}
\end{figure}

To show that our method is applicable for real-time inference, we calculated the real time factor of EEND-GLA-Small with BW-STB.
For the calculation, we used Sim5spk, in which clustering of relative speaker embeddings is always necessary (\cf \autoref{tbl:speaker_counting_simu_bw}).
The calculation was on an Intel Xeon Gold 6132 CPU @ 2.60 GHz using seven threads without any GPUs. 
Again, we used the buffer length of \SI{100}{\second} buffer and the online process unit length of \SI{1}{\second}.
\autoref{fig:rtf} shows the real time factor calculated as the processing time for each online process unit.
The real time factor increased approximately linearly until the buffer was filled, and then it became constant.
It indicates that, at least for buffer length of \SI{100}{\second}, the inference speed of EEND-GLA is not constrained by clustering of local attractors described in \autoref{sec:eend_gla}, which has $O(n^3)$ time complexity.
The convergence value of the real time factor was about 0.16 with 10 frames per second and 0.38 with 20 frames per second.
These results demonstrate that our method is fast enough for real-time inference.

\section{Conclusion}\label{sec:conclusion}
In this paper, we proposed EEND-GLA, a neural diarization method that can treat an unlimited number of speakers.
In EEND-GLA, diarization is performed on the basis of global attractors extracted from the entire input and local attractors extracted from each chunked input, respectively.
To enable online inference of EEND-GLA, we also proposed a block-wise speaker-tracing buffer; it is partitioned into blocks, and each block stores temporally continuous features and the corresponding results.
The novel speaker-balanced sampling probabilities for buffer update and variable chunk-size training via minibatch reshaping were also proposed to improve online diarization.

The experimental results showed that EEND-GLA performed well on both offline and online inferences.
In particular, EEND-GLA significantly outperformed the conventional methods on various datasets in online diarization.
The performance of the cascaded methods heavily relies on the clustering algorithm; offline diarization can utilize two-stage clustering like VBx, while online diarization cannot.
Thus, a severe gap remains between the DERs of offline and online inference of the cascaded methods.
In contrast, the offline and online DERs of EEND-GLA are less far apart than those of the cascaded methods because the inference is the same for offline and online given input features.


%

\ifCLASSOPTIONcaptionsoff
  \newpage
\fi



\bibliographystyle{IEEEtran}
%

\bibliography{mybib}

%





\end{document}